\documentclass[11pt,english]{article}
\usepackage{graphicx}
\usepackage{xcolor}
\usepackage{amstext}
\usepackage{caption}
\usepackage{etoolbox}
\usepackage{makeidx}
\usepackage{amsfonts}
\usepackage{authblk} 
\usepackage{sectsty}
\usepackage{amsmath,amssymb,epsfig}
\usepackage{amscd}
\usepackage{amsthm}
\usepackage{mathrsfs}
\usepackage{dsfont}
\usepackage[applemac]{inputenc}
\usepackage[english]{babel}
\usepackage{enumitem} 
\usepackage[]{latexsym}
\usepackage{caption}
\usepackage{subfigure}
\usepackage{hyperref}
\usepackage{blindtext}
\usepackage{verbatim}
\usepackage{cancel}
\usepackage{cases}
\usepackage{cite}

\hypersetup{
	pdftitle={},%
	pdfauthor={},%
	pdfsubject={},%
	pdfkeywords={},%
	colorlinks=true,%
	linkcolor=blue,%
	citecolor=crimson,%
	linktocpage=true,%
	hyperfootnotes=true,%
	pageanchor=true
}

\definecolor{crimson}{rgb}{0.7, 0.08, 0.24}

\newcommand{\partdif}[2]{\frac{\partial #1}{\partial #2}}

\newcommand*{\affaddr}[1]{#1}
\newcommand*{\affmark}[1][*]{\textsuperscript{#1}}

\newtheorem*{proof*}{Proof}

\newcommand{\be}{\begin{equation}}

\newcommand{\ee}{\end{equation}}
\def\beqa{\begin{eqnarray}}
\def\eeqa{\end{eqnarray}}
\def\bean{\begin{eqnarray*}}
\def\eean{\end{eqnarray*}}

\newcommand{\dd}{\mathrm{d}}

\renewcommand{\H}{\mathcal{H}}

\renewenvironment{thebibliography}[1]
         {\section*{References}\frenchspacing\small
          \begin{list}{[\arabic{enumi}]}
         {\usecounter{enumi}\parsep=2pt\topsep 0pt
         \settowidth{\labelwidth}{[#1]}
         \leftmargin=\labelwidth\advance\leftmargin\labelsep
         \rightmargin=0pt\itemsep=1pt\sloppy}}{\end{list}}

 \numberwithin{equation}{section}

\usepackage[normalem]{ulem}

\title{\textbf{\textsf{Causal Structure of a recent Loop Quantum Gravity Black Hole Collapse Model}}\vspace{0.35cm}}

\author{
\textsf{Johannes M\"unch\affmark[1]\footnote{\texttt{johannes.muench@cpt.univ-mrs.fr}}}\\
\affaddr{\affmark[1]\textsf{Aix-Marseille Université, Université de Toulon, CNRS, CPT,}}\\
\affaddr{\textsf{13288 Marseille, France}}\\
}

\begin{document}

\maketitle

\begin{abstract}
	\noindent
\textsf{The causal structure of the recent loop quantum gravity black hole collapse model \cite{KellyBlackholecollapse} is analysed.
As the spacetime is only approximately diffeomorphism invariant up to powers of $\hbar$, it is not straightforwardly possible to find global conformally compactified coordinates or to construct the Penrose diagram.
Therefore, radial ingoing and outgoing light rays are studied to extract the causal features and sketch a causal diagram.
It was found that the eternal metric \cite{KellyEffectiveloopquantumgravity}, which is the vacuum solution of the collapse model, has a causal horizon.
However, in the collapsing case light rays travel through matter to causally connect the regions inside and outside the horizon --- the causal horizon is not present in the collapsing scenario.
It is worked out that this is related to the shock wave and spacetime discontinuity, which allows matter travelling super-luminally along a space-like trajectory from the vacuum point of view, but remaining time-like from the matter perspective.
The final causal diagram is a compact patch of a Reissner-Nordström causal diagram.
Further, possibilities of a continuous matter collapse with only time-like evolution are studied. 
It is found that the time-reversed vacuum metric is also a solution of the dynamical equations and a once continuously differentiable matching of the vacuum spacetime across the minimal radius is possible.
This allows an everywhere continuous and time-like collapse process at the cost of an infinite extended causal diagram.
This solution is part of an infinitely extended eternal black hole solution with a bounce, whose global extension is constructed. 
Due to the analysis of radial light rays, it is possible to sketch causal diagrams of these spacetimes.}
\end{abstract}

\section{Introduction}

The collapse of astrophysical stars can lead under the right circumstances to the formation of a black hole.
As was shown in the singularity theorems by Hawking and Penrose \cite{PenroseGravitationalCollapseAnd,HawkingPropertiesOfExpanding}, general relativity generically predicts the formation of a spacetime singularity in this black hole collapse.
At this singularity, spacetime curvature diverges and general relativity is not further applicable.
It is widely believed that a theory of quantum gravity would allow a description of matter and spacetime at these regions \cite{BojowaldSingularitiesAndQuantum, NatsuumeTheSingularityProblem}.
Therefore, black holes have gained much attention in the context of quantum gravity.
Other fascinating puzzles like the information loss paradox and the end of black hole evaporation are motivations to study black holes in the context of quantum gravity.

One candidate for such a theory of quantum gravity is loop quantum gravity (LQG).
After techniques developed within this framework were successfully applied to cosmology, so-called loop quantum cosmology (LQC) \cite{AshtekarQuantumNatureOf,AshtekarMathematicalStructureOf,AshtekarLoopQuantumGravity30Years,DaporCosmologicalEffectiveHamiltonian} (also note recent criticism \cite{BojowaldCriticalEvaluationof}), the ground was set for efforts to understand better the quantum properties of black holes.
The essential observation here is that the black hole interior takes the form of a Kantowski-Sachs cosmological spacetime, which in principle allows extending the developed LQC techniques.
Although there has been a great amount of effort in this direction  \cite{VakiliClassicalpolymerizationof,CorichiLoopquantizationof,ModestoSemiclassicalLoopQuantum,BoehmerLoopquantumdynamics,BenAchourPolymerSchwarzschildBlack,AshtekarQuantumExtensionof,AshtekarQuantumTransfigurarationof,BodendorferEffectiveQuantumExtended,BodendorferAOSNote,BodendorferMassandHorizon,Bodendorferbvtypevariables,AssanioussiPerspectivesonthe,KellyEffectiveloopquantumgravity,KellyBlackholecollapse,GambiniSphericallysymmetricloop,GeillerSymmetriesofthe,SartiniQuantumdynamicsof,Bouhmadi-LopezAsymptoticnon-flatness,GanPropertiesofthe,Garcia-QuismondoExpolringalternativestothe,GambiniLQBHextensions}, there is no clear answer yet.
Furthermore, this framework is mainly suited for eternal black holes, where the collapse and formation process is neglected and the system reduces to an effective point mechanics problem --- so-called mini-superspace models.
The collapse process has to take the time dependence of spacetime into account.
The mini-superspace model becomes a far more complicated proper field theory, and the adaption of LQC techniques becomes more involved.
Regardless, there are efforts to understand the more complicated black hole collapse in the context of effective LQG \cite{ModestoGravitationalcollapsein,HusainCriticalbehaviourin,HossenfelderAmodelfor,BambiNon-singularquantum,BojowaldBlackHoleMass,HaggardBlackHoleFireworks,BarceloThelifetimeproblem,BianchiWhiteHolesasRemnants,Martin-DussaudEvaporationgblackto,ChakrabartyAtoymodelfor,BenAchourBouncingcompactobjectsI,BenAchourBouncingcompactobjectsII,BenAchourConsistentblackto,MuenchEffectiveQuantumDust}, but no consensus has emerged.

Very recently, there was a new effort to address the problem of a loop quantum gravity black hole \cite{KellyEffectiveloopquantumgravity} and also its formation \cite{KellyBlackholecollapse}.
Further, a full quantum theory approach was investigated independently in \cite{GambiniSphericallysymmetricloop}, which turns out to be similar to \cite{KellyEffectiveloopquantumgravity}.
It was possible to derive a regular black hole spacetime with a minimal radius instead of a singularity.
In addition, it was also possible to derive equations describing an inhomogeneous dust collapse (Lemaître-Tolman-Bondi LTB spacetimes \cite{LemaitreLuniverseen,TolmanEffectofinhomogenity,BondiSphericallySymmetricalModels}), which were solved in the homogeneous Oppenheimer-Snyder (OS) \cite{OppenheimerOnContinuedGravitational,DattOnaclass} case.
The resulting spacetime describes the formation and vanishing of black hole horizons and is complete and regular everywhere.
The collapsing matter is stopped by quantum gravity effects and re-expands outwards again, which makes the black hole disappear after some time.
This time span, the black hole lifetime, was estimated to be of order $M^2$.

This model is an interesting candidate for better understanding the life of a quantum-corrected black hole and has many physically exciting properties.
However, a global picture of the causal structure is missing in \cite{KellyBlackholecollapse,KellyEffectiveloopquantumgravity}.
Exactly this is addressed in the present paper.
An occurring obstacle is the fact that the models \cite{KellyBlackholecollapse,KellyEffectiveloopquantumgravity} are not diffeomorphism invariant, as noted by the authors.
Changing the spacetime according to a diffeomorphism leads to a different spacetime, which is corrected by contributions suppressed by Planck-order parameters.
Let us note that this seems to be a common property of this kind of models as noted in \cite{BojowaldCovarianceinmodels,BenAchourPolymerSchwarzschildBlack,BojowaldCriticalEvaluationof,BojowaldAno-goresult,BojowaldBlack-holemodels,ArrugaDeformedGeneralRelatiity}.
However, this makes it subtle to find global extensions and Kruskal-Szekeres-like coordinates, as each change of coordinates might change the physics due to the breaking of diffeomorphism invariance.
The strategy applied in this paper is therefore to keep the coordinates fixed, but study the ingoing and outgoing radial light-like geodesics, which capture the causal structure.
This is done for the vacuum spacetime region point of view and also for the matter region, which allows us to gain a global picture of the causal structure.
It is then possible to sketch a conformal diagram based on these computations.

It is shown that the collapsing matter moves super-luminally along a space-like trajectory in the re-expanding branch, which allows us to resolve the trapping horizon from the vacuum-region point of view.
Nonetheless, from the matter perspective, the matter is at rest, and is thus moving along a time-like trajectory.
These different perceptions can be traced back to a spacetime discontinuity and a shock wave during the re-expansion \cite{KellyBlackholecollapse}.
It is therefore analysed whether the underlying dynamical equations for spacetime also admit a solution where the collapse is continuous everywhere.
It turns out that this is possible and a global picture of this continuous process is derived.
The two solutions differ dramatically and differences among them and relations to previous models are discussed.

The paper is organised as follows: In Sec.~\ref{sec:model}, a review of the constructions in \cite{KellyBlackholecollapse,KellyEffectiveloopquantumgravity} is given.
Sec.~\ref{sec:causalstructure} addresses the question of the causal structure of \cite{KellyBlackholecollapse}, discussing the vacuum region in Sec.~\ref{sec:radialgeos} and the dynamics of the matter region in Sec.~\ref{sec:matter}.
The section closes by putting these results together in the causal diagram.
Finally, in Sec.~\ref{sec:discont} continuous solutions are derived and their causal structure analysed.
The paper closes with the conclusions in Sec.~\ref{sec:conclusions}.
In App.~\ref{app:fullsolutions}, the  full solutions of the eternal equations of \cite{KellyEffectiveloopquantumgravity} without fixing a gauge are presented.
This allows for an understanding of the continuous collapse of Sec.~\ref{sec:discont} from the vacuum perspective.

\section{Loop Quantum Gravity Black Hole Collapse}\label{sec:model}

The model \cite{KellyBlackholecollapse}, which is an extension of \cite{KellyEffectiveloopquantumgravity}, describes a quantum-corrected version of a spherically symmetric dust collapse (LTB spacetimes \cite{LemaitreLuniverseen,TolmanEffectofinhomogenity,BondiSphericallySymmetricalModels}).
The starting point is the spherically symmetric and time-dependent metric 
\begin{equation}\label{eq:metric}
	\dd s^2 = - N^2 \dd t^2 + \frac{\left(E^b\right)^2}{E^a} \left(\dd x + N^x \dd t\right)^2 + E^a \dd \Omega^2\;,
\end{equation}
\noindent
with $\dd\Omega^2 = \dd \theta^2 + \sin\left(\theta\right)^2\dd \phi$ being the line element of a two-sphere; $N(x,t)$ and $N^x(x,t)$ the lapse and shift; and $E^a(x,t)$, $E^b(x,t)$ the densitised triads in the angular and radial directions.
They are canonically conjugate to $a(x,t)$ and $b(x,t)$, respectively, which are the components of the spherically symmetric Ashtekar-Barbero connection 
$$
A^i_a \tau_i \dd x^a = a \tau_1 \dd x + \left(b \tau_2 - \frac{\partial_x E^a}{2 E^b} \tau_3\right) \dd \theta + \left(-\cot\left(\theta\right) \tau_1 + \frac{\partial_xE^a}{2 E^b} \tau_2 + b\tau_3\right) \sin\left(\theta\right) \dd \phi \;,
$$
\noindent
with $\tau^j = -i \sigma^j/2$ and the Pauli matrices $\sigma^j$.

The dynamics of LTB spacetimes is described classically by the Einstein-Hilbert action coupled to a dust field $T$.
In \cite{KellyBlackholecollapse}, several physically reasonable gauges are chosen at this point.
First, the clock is chosen to be the dust field itself, i.e., $t = T$; and second, the areal radius is chosen as spatial coordinate, which gives the constraint $E^a = x^2$.
The action is then completely gauge fixed and depends only on $E^b$ and $b$.
It is given by
\begin{align}
	S_{GF} =& \int\dd t \int \dd x \left(\frac{\dot{b} E^b}{G \gamma} - \H_{phys}\right)
	\\
	\H_{phys} =& -\frac{1}{2 G \gamma} \left[\frac{E^b}{\gamma x}\partial_x \left(x b^2\right) + \frac{\gamma E^b}{x} +\frac{2 \gamma x^2}{\left(E^b\right)^2} \partial_x E^b - \frac{3 \gamma x}{E^b}\right] \;,
\end{align}
\noindent
where $\gamma$ is the Barbero-Immirzi parameter, $G$ is the gravitational constant and $\H_{phys}$ is a physical Hamiltonian.
The $t= T$ dynamics is generated by $\int \dd x \H_{phys}$.
There are no remaining constraints.
The metric takes the simplified gauge fixed form
\begin{equation}
	\dd s^2 = - \dd t^2 + \frac{\left(E^b\right)^2}{x^2} \left(\dd x + N^x \dd t\right)^2 + x^2 \dd \Omega^2\;,
\end{equation}
\noindent
with $N^x = -b/\gamma$.
The only non-trivial Poisson bracket is 
\begin{equation}
	\left\{b(x_1,t),E^b(x_2,t)\right\} = G\gamma\delta\left(x_1-x_2\right) \,.
\end{equation}
\noindent
The energy density of the dust field is given by
\begin{equation}\label{eq:rhodef}
	\rho = - \frac{\H_{phys}}{4 \pi x E^b} \;,
\end{equation}
\noindent
which closes the classical theory.

Quantum effects are introduced on the effective level using a LQG argument (see \cite{KellyEffectiveloopquantumgravity,KellyBlackholecollapse} for details).
In summary, the remaining connection component $b$ is replaced by a holonomy according to
$$
b \mapsto \frac{x}{\sqrt{\Delta}} \sin\left(\frac{\sqrt{\Delta} b}{x}\right) \,,
$$
\noindent
where $\Delta \sim \ell_p^2$ is the smallest non-zero eigenvalue of the area operator in LQG.
The resulting Hamiltonian is
\begin{equation}\label{eq:Heff}
	\H_{phys}^{eff} = -\frac{1}{2 G \gamma} \left[\frac{E^b}{\gamma x}\partial_x \left(\frac{x^3}{\Delta} \sin\left(\frac{\sqrt{\Delta} b}{x}\right)^2\right) + \frac{\gamma E^b}{x} +\frac{2 \gamma x^2}{\left(E^b\right)^2} \partial_x E^b - \frac{3 \gamma x}{E^b}\right] \;,
\end{equation}
\noindent
while the Poisson brackets remain unchanged.
The relation of the canonical variables to the lapse $N^x$ is modified according to \cite{KellyEffectiveloopquantumgravity,GambiniSphericallysymmetricloop}
\begin{equation}\label{eq:Nx}
	N^x = -\frac{x}{\gamma \sqrt{\Delta}} \sin\left(\frac{\sqrt{\Delta} b}{x}\right)\cos\left(\frac{\sqrt{\Delta} b}{x}\right) \;.
\end{equation}

The quantum modified equations of motion are finally given by
\begin{subequations}\label{eq:EoM}
	\begin{align}
		\dot{E}^b =& - \frac{x^2}{2 \gamma \sqrt{\Delta}} \partial_x \left(\frac{E^b}{x}\right) \sin\left(\frac{\sqrt{\Delta} b}{x}\right)\cos\left(\frac{\sqrt{\Delta} b}{x}\right)\label{eq:dotEb} \;,
		\\
		\dot{b} =& \, \frac{\gamma}{2} \left(\frac{x}{\left(E^b\right)^2} - \frac{1}{x}\right) - \frac{1}{2 \gamma \Delta x} \partial_x \left(\frac{x^3}{\Delta} \sin\left(\frac{\sqrt{\Delta} b}{x}\right)^2\right) \;,
	\end{align}
\end{subequations}
together with the relation to the dust field Eq.~\eqref{eq:rhodef}. 

These equations are solved in \cite{KellyBlackholecollapse} for the simplest LTB case of a homogeneous Oppenheimer-Synder dust collapse \cite{OppenheimerOnContinuedGravitational,DattOnaclass}.
The dust is therefore assumed to be homogeneous in the spacetime region up to an areal radius $L(t)$ and zero otherwise.
Eq.~\eqref{eq:dotEb} is obviously solved for $E^b = x$, leaving two remaining equations

\begin{align}
	\dot{b} =& -\frac{1}{2\gamma \Delta x} \partdif{}{x} \left(x^3 \sin^2\left(\frac{\sqrt{\Delta} b}{x}\right)\right)\;,
	\label{eq:b}
	\\
	\rho =& \frac{1}{8 \pi G \gamma^2 \Delta x^2} \partdif{}{x} \left(x^3 \sin^2\left(\frac{\sqrt{\Delta} b}{x}\right)\right) \;.\label{eq:rho}
\end{align}
\noindent
The equations decouple then into the region with $\rho \neq 0$ and $\rho = 0$ and can be solved interdependently.
An important assumption for this being true is the neglect of edge effects (see \cite{KellyBlackholecollapse}).
Following the steps in \cite{KellyBlackholecollapse} leads to the complete dynamical spacetime, which is given by

\begin{equation}
	\dd s^2 = -\left(1-\left(N^x(x,t)\right)^2\right) \dd t^2 +2 N^x(x,t)\dd t \dd x + \dd x^2 + x^2 \dd \Omega^2 \;,
\end{equation}
\noindent
with $N^x = N_{mat}^x$ for $x < L(t)$ and $N^x = N^x_{vac}$ for $x>L(t)$, given by
\begin{subequations}
	\begin{align}
		N^x_{mat}(x,t) =& -\frac{\dot{L}}{L} x
		\quad ,\quad  x < L(t) 
		\;,\label{eq:dsint}
		\\
		N^x_{vac}(x) =& \sqrt{\frac{R_s}{x}\left(1-\frac{\gamma^2 \Delta R_s}{x^3}\right)} 
		\quad ,\quad  x > L(t) 
		\;.\label{eq:dsext}
	\end{align}	
\end{subequations}
\noindent
Here $R_s = 2 G M$, with the mass of the black hole $M$.
The dynamics of the outermost dust shell is given by\footnote{This expression is assumed to hold only in the absence of edge effects. As stated in \cite{KellyBlackholecollapse} edge effects are relevant in the re-expansing branch, where the metric is discontinuous and the shock wave develops. As also noted in \cite{KellyBlackholecollapse} edge effects might not cure the discontinuity, but eventually they change the expansion rate and the exact values in the equation $L(t)$ for $t > 0$.}
\begin{equation}\label{eq:Lsol}
	L(t) = \left[x_{min}^3 \left(\frac{9 t^2}{4 \gamma^2 \Delta}+1\right)\right]^\frac{1}{3} \;,
\end{equation}
\noindent
with $x_{min} = \left(\gamma^2 \Delta R_s\right)^{\frac{1}{3}}$.
Note that $L(t)$ satisfies

\begin{equation}\label{eq:FriedmannEq}
	\left(\frac{\dot{L}}{L}\right)^2 = \frac{8\pi G}{3} \rho \left(1-\frac{\rho}{\rho_c}\right) = \frac{R_s}{L^3} \left(1-\frac{x_{min}^3}{L^3}\right) = \frac{\left(N_{vac}^x\right)^2}{L^2} \;,
\end{equation}
\noindent
with $\rho = 3M/4\pi L^3$ and $\rho_c = 3 R_s/8\pi G x_{min}^3$, which is simply the LQC effective equation. 
The vacuum region of the spacetime was independently found in \cite{KellyEffectiveloopquantumgravity}.

The spacetime consists of a shock wave \cite{KellyBlackholecollapse}, in which the matter does re-expand, i.e. $t > 0$.
This causes a discontinuity in the metric across the surface $x = L(t)$, as can be straightforwardly seen:
\begin{subequations}
	\begin{align}
		\left.\dd s_{mat}^2\right|_{x = L(t)} =& -\dd t^2 + L(t)^2 \dd \Omega^2 \;,
		\\
		\left.\dd s_{vac}^2 \right|_{x = L(t)} =&  \begin{Bmatrix}
			-1 &,\; t < 0
			\\
			-1+4 \dot{L}^2 &,\; t > 0
		\end{Bmatrix} \dd t^2 + L(t)^2 \dd \Omega^2 \;.
	\end{align}
\end{subequations}
\noindent
The induced metrics match obviously for $t<0$, but not for the re-expansion.
The reason for this is that $N^x_{vac} > 0$ for all times, while $N_{mat}^x$ changes its sign across the bounce.

\section{Causal Structure}\label{sec:causalstructure}

We would like to analyse the causal structure of this LQG black hole collapse model.
The main obstacle here is the fact that the model is not classically diffeomorphism invariant as was noted by the authors of \cite{KellyEffectiveloopquantumgravity}.
The symmetry is violated, by corrections suppressed by Planck-order contributions, though.
Therefore, it is not possible to find Kruskal-Szekeres-like global coordinates and perform a conformal compactification as this would change the spacetime.
As shown in \cite{KellyEffectiveloopquantumgravity}, if the fundamental equations are solved in different coordinates, i.e. using different lapse functions $N$ leads to physically different spacetimes (see also App.~\ref{app:fullsolutions})\footnote{As shown in App.~\ref{app:fullsolutions}, choosing the lapse appropriately leads to the same metric as in \cite{GambiniSphericallysymmetricloop}, neglecting discreteness effects. However, \cite{GambiniSphericallysymmetricloop} derives this metric from a quantum theory rather than being an effective level consideration and contains further corrections. The two models share similar features; e.g. the dependence on the choice of lapse and shift, and thus the analysis of this section, can be straightforwardly applied to the effective metric \cite{GambiniSphericallysymmetricloop}, and similar results are expected. Nevertheless, here the collapse process as in \cite{KellyBlackholecollapse} is not included. The causal structure of \cite{GambiniSphericallysymmetricloop} needs a separate examination using the same techniques as discussed in Sec.~\ref{sec:radialgeos}. As the main interest of the present paper is the understanding of the global structure of the collapse and \cite{GambiniSphericallysymmetricloop} does not include it, this analysis is reported elsewhere.}.
However, the model is valid as long as we use the same Painlevé-Gullstrand-like coordinates.
It is then possible to simply study light-like radial geodesics to extract the causal structure.
It is not clear if this is really compatible with the quantum diffeomorphisms braking, but on the effective level, it is a minimal requirement for well defining a causal structure and is in fact a valid spacetime description.
In general, the discussion on diffeomorphism invariance at the quantum level is ongoing \cite{BojowaldCovarianceinmodels,BenAchourPolymerSchwarzschildBlack,BojowaldCriticalEvaluationof,BojowaldAno-goresult,BojowaldBlack-holemodels,ArrugaDeformedGeneralRelatiity}.

\subsection{Asymptotic Structure and Horizons}

Let us first recall the main features of the vacuum-region spacetime, i.e. $x > L(t)$.
In the limit $x \gg x_{min}$ and $x > L(t)$, the metric simplifies to 

\begin{equation}
	N_{vac}^x \simeq \sqrt{\frac{R_s}{x}}\left(1 + \mathcal{O}\left(\frac{x_{min}^3}{x^3}\right)\right) \;,
\end{equation}
\noindent
and thus the asymptotic vacuum spacetime

\begin{equation}
	\dd s_{vac}^2 \simeq - \left(1-\frac{R_s}{x}\right) \dd t^2 + 2 \sqrt{\frac{R_s}{x}} \dd t \dd x + x^2 \dd\Omega^2 + \mathcal{O}\left(\frac{x_{min}^3}{x^3}\right)\;,
\end{equation}
\noindent
becomes the Schwarzschild metric in Painlevé-Gullstrand coordinates.
As the spacetime is asymptotically Schwarzschild, it is clear that it is asymptotically flat for $x/R_s \gg 1$.

Let us note that there are two horizons, which would correspond to Killing horizons (ignoring the matter) in the static case, i.e. $g(\xi_t,\xi_t) = 0$ with $\xi = \partial/\partial t$.
These are given by
\begin{equation}
	N_{vac}^x(x_{in/out}) = 1 \quad , \quad x_{out} \simeq R_s \quad , \quad x_{in} \simeq x_{min} \;.
\end{equation}
\noindent
Note that for both solutions of $-g_{tt} = 1-\left(N_{vac}^x\right)^2 = 0$, the shift is positive $N_{vac}^x(x_{in/out}) > 0$.
Obviously there is also the solution $N_{vac}^x(x) = -1$, but this is never reached as $N_{vac}^x(x)$ is defined as a positive square root, see Eq.~\eqref{eq:dsext}.
This equation can be written out to give
\begin{equation}\label{eq:horizonrewritten}
	1-\frac{R_s}{x} \left(1-\frac{x_{min}^3}{x^3}\right) = 0 \quad \Leftrightarrow \quad \frac{R_s x_{min}^3}{x^4} = \frac{R_s}{x} -1 \;.
\end{equation}
\noindent
The second re-writing of the equation will later be of use.
As this equation is a fourth-order polynomial, it is possible to construct analytic solutions, but their expression will be very complicated.
We see immediately that this equation is solved for $x_{out} \simeq R_s \left(1- \mathcal{O}\left(x_{min}^3/R_s^3\right)\right)$, and also for $x_{in} \simeq x_{min}\left(1 + \mathcal{O}\left(x_{min}/R_s\right)\right)$ (cfr. \cite{KellyEffectiveloopquantumgravity}).

The notion of a Killing horizon is not present in the present dynamical setting, where dynamical matter is included and the Killing symmetry does not exist globally.
A better suited notion is therefore a trapping horizon, also called an apparent horizon.
These horizons are defined as the boundaries of trapped regions, and in a static setting the notions of Killing and trapping horizons coincide.
As was computed in \cite{KellyEffectiveloopquantumgravity}, such horizons are indeed apparent horizons, as the expansions are given by

\begin{equation}
	\theta_+ = \frac{2}{x} \left(1-N_{vac}^x\right) \quad , \quad \theta_- = -\frac{2}{x} \left(1-N_{vac}^x\right) \;.
\end{equation}
\noindent
Fig.~\ref{fig:psi} (a) shows a plot of $N_{vac}^x$, which makes it clear that both expansions have negative sign in the region $x_{in} < x < x_{out}$, while they have different signs for $x>x_{out}$ and $x<x_{in}$.
Consequently, the region $x_{in} < x < x_{out}$ is trapped and as $x = x_{in}$ and $x_{out}$ are the boundaries of this region, these hyper-surfaces are apparent or trapping horizons.

\subsection{Radial Light-like Geodesics in the Vacuum Region}\label{sec:radialgeos}

Let us now study radial light-like geodesics.
At first, we ignore the presence of matter, which places us in the setup of \cite{KellyEffectiveloopquantumgravity}.
Nevertheless, we should keep in mind that there is matter in this spacetime.
The correct treatment considers that if the geodesic evolves into $x = L(t)$ i.e., it hits the matter surface the continuing evolution is described by the geodesic equation following from the matter metric~\eqref{eq:dsint}.
Exactly this is ignored for the geodesic evolution in this section, but it is corrected in the following section.
Still, we will comment on the influence of matter.

The main question arising is, whether these apparent horizons of the previous sections are also causal horizons i.e., if there exist light rays, which can still reach $x \rightarrow \infty$ once the horizon is passed.
To address this question, we can solve the light-like radial ingoing and outgoing geodesic equations in the exterior i.e., assuming that the light was emitted at any point $(t,x)$ with $x > L(t)$.
Radial light-like geodesics satisfy
\begin{equation}
\dd s^2 = 0 = - \left(1-\left(N_{vac}^x(x)\right)^2\right) \dd t^2 + 2N_{vac}^x(x) \dd t \dd x \;,
\end{equation} 
\noindent
thus giving the differential equations 
\begin{equation}
	\frac{\dd x}{\dd t} = -N_{vac}^x(x) \pm 1 \;,
\end{equation}
\noindent
where the $+$ sign corresponds to outgoing radial geodesics, and the $-$ sign to ingoing ones.
\begin{figure}[h!]
	\centering
	\subfigure[]
	{\includegraphics[width=7.3cm]{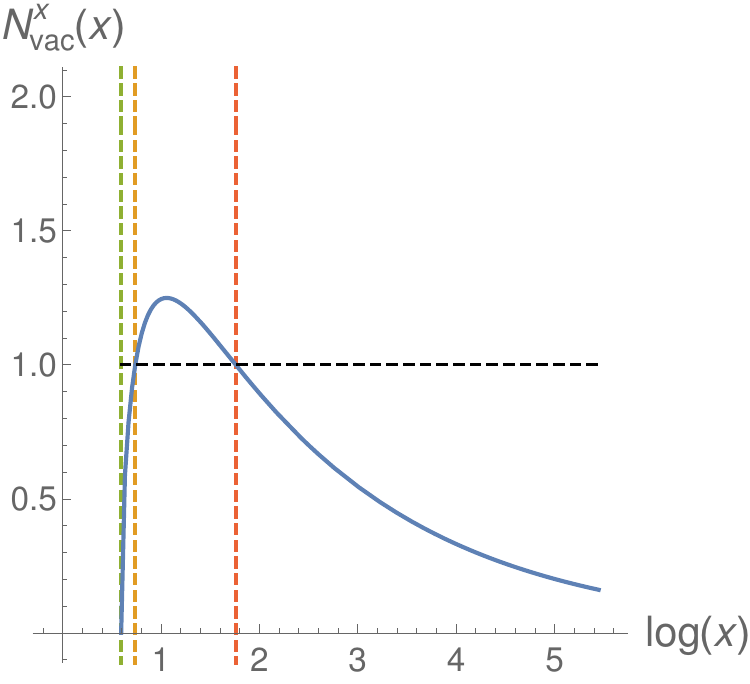}}
	\hspace{2mm}
	\subfigure[]
	{\includegraphics[width=7.3cm]{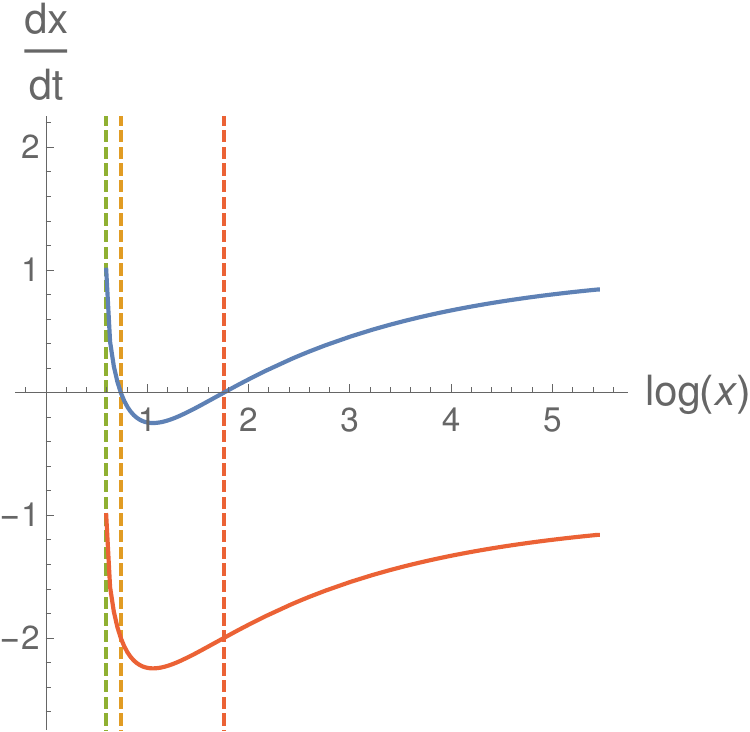}}
	\caption{(a) Plot of the function $N^x_{vac}(x)$ and (b) the phase-space plot $\frac{\dd x}{\dd t}$ vs $x$ for both ingoing (red) and outgoing (blue) light-like geodesics. Green, yellow and red dashed vertical lines correspond to $x_{min}$, $x_{in}$, and $x_{out}$, respectively. The black dashed line in (a) indicates $N_{vac}^x(x) = 1$; i.e., where $\dd x^+/\dd t = 1- N_{vac}^x(x) < 0$. The parameters are chosen as $R_s = 6$, $\gamma = \Delta = 1$.}
	\label{fig:psi}
\end{figure}
Consistently, we have for $x>x_{out}$ that $\dd x / \dd t = -N_{vac}^x +1 > 0$.
It is possible to draw a phase-space plot for the outgoing geodesics; see Fig.~\ref{fig:psi}.
From the plot, it is evident that 

\begin{equation}
	\frac{\dd x^+}{\dd t} := -N_{vac}^x(x) + 1 = \begin{cases}
		> 0 &\,,\; x > x_{out} \\
		< 0 &\,,\; x_{in} < x < x_{out}
	\end{cases} \,,
\end{equation}
\noindent
while it is indefinite for $x < x_{min}$.
This is no problem, as there is no instance of time for which $L(t) < x_{min}$.

When staying outside the outer horizon, $\dd x^+/\dd t > 0$ i.e., light rays will always try to escape to infinity.
They will also reach infinity unless they first hit the surface $(t, L(t))$.

The geodesic equation cannot be solved analytically.
However, it is instructive to perform a stability analysis to determine the first-order evolution close to the horizon. 
Consider a light ray starting very close to the horizon i.e., $x(t) = x_{out} + \epsilon(t)$, with $\epsilon(t) \ll 1$.
The geodesic equation reads in this case

\begin{align}
\frac{\dd x^+}{\dd t} = \frac{\dd \epsilon}{\dd t} =&\, 1-N_{vac}^x\left(x_{out} + \epsilon(t)\right) \simeq 1- N_{vac}^x(x_{out}) - N_{vac}^{x'}(x_{out})\epsilon(t) + \mathcal{O}\left(\epsilon^2\right)
\notag
\\
=& - N_{vac}^{x'}(x_{out})\epsilon(t) + \mathcal{O}\left(\epsilon^2\right) \;,
\label{eq:approxout}
\end{align}
\noindent
as $N_{vac}^x(x_{out}) = 1$.
We abbreviate $\left.\frac{\dd N_{vac}^x}{\dd x}\right|_{x=x_{out}} = N_{vac}^{x'}(x_{out})$.
Evaluating the derivative of $N_{vac}^x$ leads to
\begin{equation}\label{eq:psiprime}
	N_{vac}^{x'}(x) = -\frac{1}{2xN_{vac}^x}\left(\left(N_{vac}^x\right)^2 -\frac{3R_s x_{min}^3}{x^4}\right)\stackrel{x = x_{out}}{=} -\frac{1}{2x_{out}}\left(1 -\frac{3R_s x_{min}^3}{x_{out}^4}\right)\;.
\end{equation}
\noindent
Using Eq.~\eqref{eq:horizonrewritten} this gives
\begin{equation}
	N_{vac}^{x'}(x_{out}) = -\frac{1}{2x_{out}} \left(1-3 \left(\frac{R_s}{x_{out}}-1\right)\right) < 0 \;,
\end{equation}
\noindent
which is always negative, as $x_{out}/R_s = 1-\mathcal{O}\left(x_{min}^3/R_s^3\right)$.
We can therefore write

\begin{equation}
	N_{vac}^{x'}(x_{out}) = -\kappa_o \quad , \quad \kappa_o = \frac{1}{2x_{out}} \left(1-3 \left(\frac{R_s}{x_{out}}-1\right)\right) > 0 \;.
\end{equation}
\noindent
The approximate geodesic equation~\eqref{eq:approxout} in the outer vicinity of the outer horizon is therefore solved by
\begin{equation}
\epsilon(t) \simeq \epsilon_0 e^{\kappa_o t} \;.	
\end{equation} 
\noindent
For large $t$ this grows exponentially, which makes the approximation invalid already after a short period of time.
The light-like geodesic, therefore, escapes quickly to larger radii where it can in principle escape to infinity.
On the other hand, for negative $t$, it approaches the horizon only asymptotically.

We can repeat the above computation for $x(t) = x_{out} - \epsilon(t)$, i.e. for a position slightly behind the outer horizon.
In this case, we find 
\begin{equation}
\frac{\dd x^+}{\dd t} = -\frac{\dd \epsilon}{\dd t} = - N_{vac}^{x'}(x_{out}) \epsilon(t) + \mathcal{O}\left(\epsilon^2\right) \,,
\end{equation}
\noindent
which is equivalently re-written as
\begin{equation}
	\frac{\dd \epsilon}{\dd t} = -\kappa_o < 0 \quad \Rightarrow \quad \epsilon(t) \simeq \epsilon_0 e^{-\kappa_o t} \;.
\end{equation}
\noindent
This shows that the geodesic behind the outer horizon exponentially grows towards smaller $x$-values, where the approximation of $\epsilon \ll 1$ is quickly broken.
Checking the phase-space plot Fig.~\ref{fig:psi}, the light ray continues to move towards smaller radii until the inner horizon is reached.
This shows, that once it is behind the outer horizon, there is no way to exit it any more from a purely exterior point of view.
Already at this stage, we see a causal structure, which is similar to a charged Reissner-Nordström black hole.
Let us also note that for $t \rightarrow -\infty$ the outer horizon is again approached exponentially.

To gain a full picture of the causal structure, the above steps are repeated for the inner horizon.
We assume $x(t) = x_{in} + \epsilon(t)$, i.e. the light ray is emitted in the trapped region between the horizons, but close to the inner horizon.
Similarly to before, we find
\begin{equation}
\frac{\dd \epsilon}{\dd t} \simeq - N_{vac}^{x'}(x_{in}) \epsilon + \mathcal{O}\left(\epsilon\right) \;.
\end{equation}
\noindent
Evaluating the derivative (cfr. Eq.~\eqref{eq:psiprime}) yields
\begin{equation}
	N_{vac}^{x'}(x_{in}) = -\frac{1}{2 x_{in}} \left(1-\frac{3 R_s x_{min}^3}{x_{in}^4}\right) = -\frac{1}{2 x_{in}} \left(4-3\frac{R_s}{x_{in}}\right) =: -\kappa_i \;,
\end{equation}
\noindent
where again Eq.~\eqref{eq:horizonrewritten} is used in the last step.
As $R_s/x_{in} \gg 1$, $\kappa_i$ is always negative.
The approximate solution is thus 
\begin{equation}
	\epsilon(t) \simeq \epsilon_0 e^{|\kappa_i| t} \;,
\end{equation}
\noindent
which shows that for positive $t$ the inner horizon is approached exponentially, while in the past ($t < 0$) the geodesic tends exponentially to larger radii.
From the phase-space plot in Fig.~\ref{fig:psi} it is evident that the light ray came from $x_{out}$.

Analogous to before, we find for $x(t) = x_{in} - \epsilon$ the solution
\begin{equation}
	\epsilon(t) \simeq \epsilon_0 e^{-|\kappa_i| t} \;.
\end{equation}
\noindent
Therefore in the future light rays approach the inner horizon exponentially and in the past, they tend to smaller radii, i.e. $x_{min}$ as can again be read from Fig.~\ref{fig:psi} (b).

This closes the stability analysis of the outgoing light rays.
In summary, $x_{out}$ is an unstable fixed point of the outgoing radial geodesic equation, while $x_{in}$ is a stable fixed point.

The analysis of the ingoing light rays becomes trivial.
As $N_{vac}^x(x) > 0$, for all $x$, we find $\dd x^-/\dd t = -N_{vac}^x(x) - 1 < 0$ everywhere.
Therefore, ingoing light rays are simply moving towards $x_{min}$.
The horizons are crossed in finite coordinate time $t$.

\begin{figure}[h!]
	\centering
	\subfigure[]
	{\includegraphics[width=7.3cm]{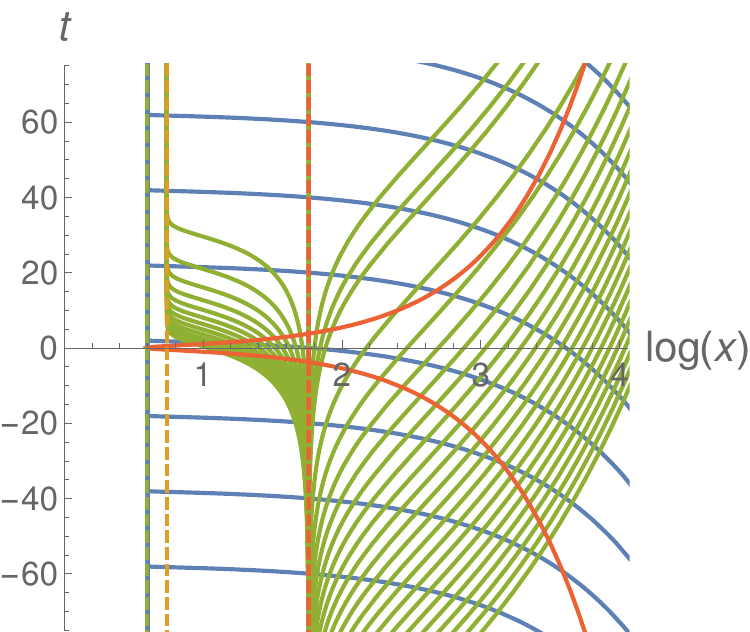}}
	\hspace{2mm}
	\subfigure[]
	{\includegraphics[width=7.3cm]{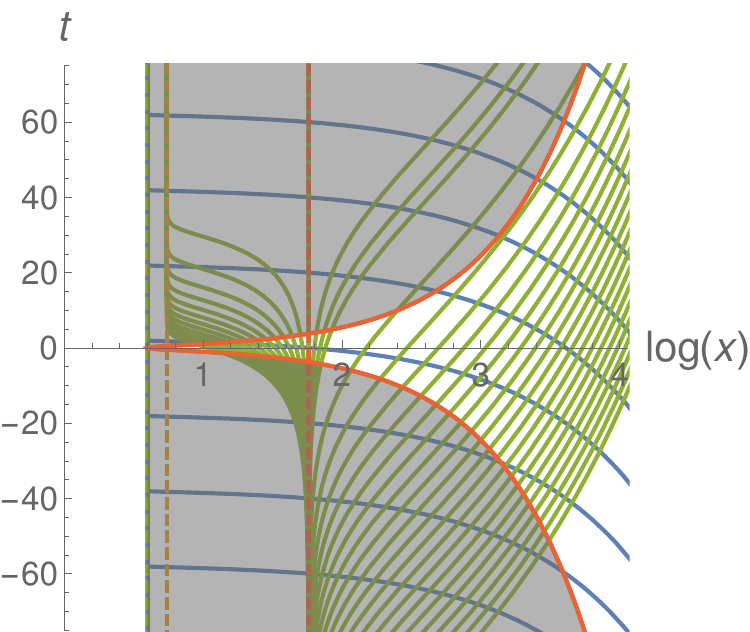}}
	\caption{(a) Plot of ingoing and outgoing radial light-like geodesics. (b) The same, with the region that has to be replaced by matter shaded grey. Blue curves are ingoing light rays, which travel in finite coordinate time to $x_{min}$. Green curves are outgoing light rays, showing nicely the features worked out in the main text. The red curve is the surface of the matter $L(t)$. The vertical lines are $x_{min}$, $x_{in}$ and $x_{out}$. Parameters are again $R_s = 6$, $\gamma = \Delta = 1$.}
	\label{fig:causalstructure}
\end{figure}

We can also solve the geodesic equations fully numerically.
The above analysis then supports the numerical results in the regions close to the horizons, where numerical uncertainties are large.
Plotting the solutions for several initial conditions in the $t-x$-chart leads to Fig.~\ref{fig:causalstructure}.
The qualitative features worked out analytically are nicely visible in this plot.
Geodesics staring\footnote{Solving the outgoing equations for $x_{min}< x <x_{in}$ is numerically more challenging as $x_{min} \sim x_{in}$ for the chosen parameters. The geodesics can nonetheless qualitatively be well understood from the analytical discussion above.} outside the outer horizon tend to infinity, while those behind the outer horizon approach either $x_{in}$.

Completely neglecting the matter leads finally to the conclusion that $x_{out}$ is a causal horizon.
There are no light-like geodesics that can start behind this horizon and still escape to infinity.
This result is rather trivial as neglecting the matter, the spacetime is static where usually all these notions of horizons coincide.
However, the spacetime in fact is dynamic and contains matter.
It might be possible that a light ray hits the matter surface, travels inside the matter and then exits the matter again at $x > x_{out}$.
In this case the region $x < x_{out}$ is actually not causally disconnected from $x > x_{out}$ and there is no causal horizon.
This might be possible, as we see from Eq.~\eqref{eq:Lsol} that $L(t\rightarrow\infty) \rightarrow \infty$ and thus becomes arbitrarily large at large times.
However, it the important question remains of how the matter surface itself can exit the outer horizon if not even light is able to do so.

\subsection{Matter Region}\label{sec:matter}

So far we have ignored the matter part of the metric and the fact that there are dynamical boundary conditions.
Exactly this is addressed in this subsection.
As just discussed light cannot exit the region $x < x_{out}$ via a geodesic lying purely in $x > L(t)$.
It still could be that a light ray exits this region by moving through the matter spacetime.
As argued, this might be possible, as $L(t)$ again becomes arbitrarily large after a bounce.
Usually, this happens in a parallel universe \cite{BenAchourBouncingcompactobjectsI,BenAchourBouncingcompactobjectsII,BenAchourConsistentblackto,MuenchEffectiveQuantumDust,SchmitzExteriorstobouncing}.

The surface of matter evolves according to Eq.~\eqref{eq:FriedmannEq}, which can equivalently be written as

\begin{equation}\label{eq:Ldotpsi}
	\dot{L}(t) = \begin{cases}
		+N_{vac}^x(x=L(t)) &,\; \text{expansion} \\
		-N_{vac}^x(x=L(t)) &,\; \text{collapse}
	\end{cases}\;.
\end{equation}
\noindent
The solution $L(t)$ is given by Eq.~\eqref{eq:Lsol}.
The spacetime trajectory of the matter surface is thus given by $\gamma = \left(t, L(t), \theta_0, \phi_0\right)$, where $\phi_0$ and $\theta_0$ indicate an arbitrary point on $\mathbb{S}^2$, which is irrelevant for the following.
Consider now an observer, sitting at $x = x_{obs} > x_{out}$ at $t = -t_{obs}$ on the matter surface.
Obviously, $x_{obs} = L(-t_{obs})$ is required to satisfy this condition.
The observer stays at this radius and watches the matter collapsing.
We can now ask: Does he touch the matter surface ever again?
To answer this question, we need to solve the equation
\begin{align}
	L(t) &= x_{obs} \quad , \quad t \in \left[-t_{obs}, \infty\right) \;,
	\notag
	\\
	\left[x_{min}^3 \left(\frac{9 t^2}{4 \gamma^2 \Delta}+1\right)\right]^\frac{1}{3} &= x_{obs} \;.
\end{align}
\noindent
As we define $t_{obs}$ by the condition $L(-t_{obs}) = x_{obs}$, this is certainly one solution of this equation.
As further $L(-t) = L(t)$, the value $t_{obs}$ also satisfies this equation, which then gives the second and last possible solutions.
For completeness, it is
$$
t_{obs} = \frac{2\gamma \sqrt{\Delta}}{3} \sqrt{\frac{x_{obs}^3}{x_{min}^3} -1 } = \frac{2\gamma \sqrt{\Delta}}{3} \sqrt{\frac{x_{obs}^3}{\gamma^2 \Delta R_s} -1 }\;.
$$
\noindent
Thus the observer meets the matter surface again after the finite time $2t_{obs}$, which translates into a finite observer eigentime $\tau = 2t_{obs}\sqrt{1-\left(N_{vac}^x(x_{obs})\right)^2}$.
As the observer and matter meet each other after finite time, this is different from most other proposals for loop quantum gravity-inspired black hole models where a causal horizon is present \cite{VakiliClassicalpolymerizationof,CorichiLoopquantizationof,ModestoSemiclassicalLoopQuantum,BoehmerLoopquantumdynamics,BenAchourPolymerSchwarzschildBlack,AshtekarQuantumExtensionof,AshtekarQuantumTransfigurarationof,BodendorferEffectiveQuantumExtended,BodendorferMassandHorizon,Bodendorferbvtypevariables,AssanioussiPerspectivesonthe,KellyEffectiveloopquantumgravity,KellyBlackholecollapse,GambiniSphericallysymmetricloop,GeillerSymmetriesofthe,SartiniQuantumdynamicsof,SchmitzExteriorstobouncing,GanPropertiesofthe}.
A generalisation to a collapse model as in \cite{BenAchourBouncingcompactobjectsI,BenAchourBouncingcompactobjectsII,BenAchourConsistentblackto,MuenchEffectiveQuantumDust,SchmitzExteriorstobouncing} would always lead to the result that the observer and matter never meet again as the matter bounces out of a parallel universe --- at least if black hole evaporation is neglected.
Here instead, the matter is crossing the outer horizon, then the inner one, bouncing out again at $x_{min}$, crossing first the same inner horizon and then the same outer horizon to arrive in the \textit{same exterior} again.
On the one hand, this gives the opportunity for a causal horizon to be avoided, as light rays could enter the matter region and exit the matter once it has again passed $x_{out}$.
On the other hand, this raises the question of how this is possible when we have argued that $x_{out}$ causally disconnects the inner and outer spacetime regions from a vacuum point of view.
The only possible solution is that the matter moves super-luminally along a space-like trajectory in some time interval.
This way it would be possible to come out of the outer horizon again.

\begin{figure}[h!]
	\centering\includegraphics[width=7.35cm]{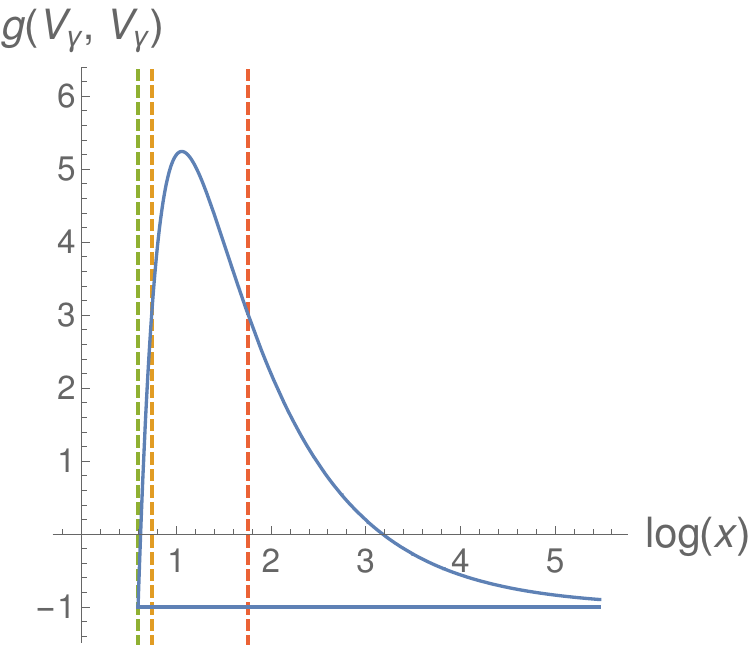}
	\caption{Plot of $g\left(V_\gamma,V_\gamma\right)$ against the position $x$. It is clearly visible that on the collapsing branch, the norm of the tangent is constant at $-1$. On the re-expanding branch, the norm changes in time and becomes space-like in between the horizons, and even stays space-like for some time once the outer horizon is again passed. Dashed lines indicate (from left to right) $x_{min}$, $x_{in}$, and $x_{out}$. The parameters are $R_s = 6$, $\gamma = \Delta = 1$.}
	\label{fig:gVV}
\end{figure}

It is easy to check this property.
First, the tangent to the trajectory $\gamma$ is given by
\begin{equation}
	V_{\gamma} = \partdif{}{t} + \dot{L} \partdif{}{x} \;.
\end{equation}
\noindent
Its norm allows us to track the causality of this curve.
Consequently, it is
\begin{align}
	g\left(V_\gamma,V_\gamma\right) &= -\left(1-\left(N_{vac}^x(L(t))\right)^2\right) + 2 N_{vac}^x(L(t)) \dot{L}(t) + \dot{L}(t)^2 
	\notag
	\\
	&= -1 +2 \left( \text{sign}\left(\dot{L}\right)+1\right)\left(N_{vac}^x(L(t))\right)^2 = \begin{cases}
		-1 &,\; \dot{L} < 0 \\
		-1 + 4 \dot{L}^2 &, \; \dot{L}> 0
	\end{cases}\;.
\end{align}
\noindent
Here we use the equation for $\dot{L}$ in the form of Eq.~\eqref{eq:Ldotpsi}.
Obviously, in the collapsing branch, i.e., $\dot{L} < 0$, this is constant at $-1$, i.e., the tangent is normalised and time-like for the full collapsing branch.
This is not the case for the re-expanding branch, i.e. $\dot{L} > 0$.
We know that at the horizons $N_{vac}^x(x_{in/out})  = 1$ and further $N_{vac}^x(x) > 1$ for $x \in \left(x_{in},x_{out}\right)$.
Therefore, the tangent is space-like at least within the inner and outer horizons as there
$$
g\left(V_\gamma,V_\gamma\right) = -1 + 4 \left(N_{vac}^x\right)^2 > 3 \quad ,\quad \text{for } x_{in} < L(t) < x_{out} \;.
$$ 
\noindent
As $N_{vac}^x(x\rightarrow\infty) \rightarrow 0$, this becomes again time-like far enough from the outer horizon.
The full plot of $g\left(V_\gamma,V_\gamma\right)$ is shown in Fig.~\ref{fig:gVV}, which verifies this result.

This spacetime picture concerns only the point of view of an exterior observer and how light rays propagate in the vacuum part of the metric.
Let us emphasise that the super-luminal motion of the matter surface is not present from the matter point of view.
It is easy to compute the norm of the tangent for the metric~\eqref{eq:dsint}.
This simply leads to
\begin{equation}
	g\left(V_\gamma,V_\gamma\right) = -\left(1-\left(N_{mat}^x\right)^2\right) +2 N_{mat}^x \dot{L}+ \dot{L}^2 = -\left(1-\dot{L}^2\right) -2 \dot{L}^2+ \dot{L}^2 = -1 \;.
\end{equation}
\noindent
Therefore, for all times, the tangents remain time-like w.r.t. the interior point of view.
This is only possible due to the spacetime discontinuity caused by the shock wave.

It remains the question if light can actually leave the matter region once it is caught by it.
From the matter metric point of view, the outgoing light-like geodesic equation reads
\begin{equation}
	\frac{\dd x^+}{\dd t} = -N_{mat}^x +1 = \frac{\dot{L}}{L} x + 1 \quad , \quad \frac{1}{x} \frac{\dd x}{\dd t} = \frac{\dot{L}}{L} + \frac{1}{x} > \frac{\dot{L}}{L} \;. 
\end{equation}
\noindent
This implies that first of all, $\frac{\dd x}{\dd t}$ is always positive, i.e. light moves really outwards and second, it moves faster than the outermost matter.
Thus the surface of the matter is always reached sooner or later.
In conclusion, light can enter the matter region at some point $x < x_{min}$ moves outward with the matter and exits at $x > x_{out}$.
Therefore, the static setting of Sec.~\ref{sec:radialgeos} is miss-leading and taking the dynamical matter into account, the regions $x > x_{out}$ and $x< x_{out}$ are not causally disconnected and there is no causal horizon in the spacetime.  

\begin{figure}[h!]
	\centering\includegraphics[width=9cm]{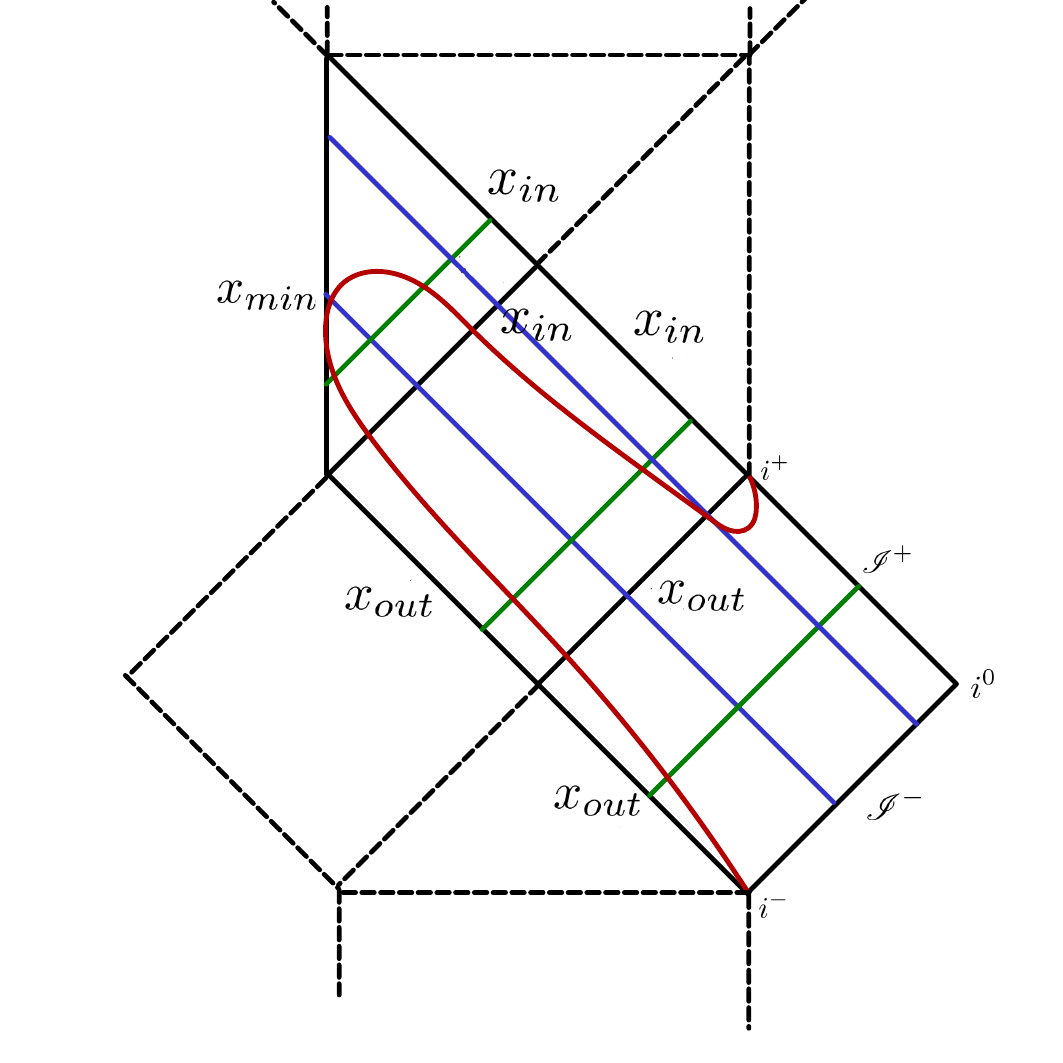}
	\caption{Schematic Penrose diagram. Here, ingoing (blue) and outgoing (green) light rays are drawn orthogonally to each other. This summarises all the above derived features qualitatively. The red curve is a sketch of $L(t)$, which nicely cuts out the all non-wanted spacetime regions, as the vacuum spacetime is only valid up to this point. 
	Behind it the matter metric has to be used. The Painlevé-Gullstrand coordinates only cover the solid black region.}
	\label{fig:penrose}
\end{figure}

Having all this analysis done, it is possible to construct a causal diagram.
First, the plot of ingoing and outgoing light-like geodesics as well the surface of the matter, is given in Fig.~\ref{fig:causalstructure}.
It shows nicely the discussed features.
All light rays behind the outer horizon are caught again by the matter surface.
This is clear, as in the infinite future and past they approach either the inner or outer horizon.
This cures the missing description of the exterior spacetime at radii $x < x_{min}$ as already discussed in \cite{KellyBlackholecollapse,KellyEffectiveloopquantumgravity}.
Outside the outer horizon, some of the outgoing light rays are caught by the matter surface, while others are not.
Note that asymptotically ($t \rightarrow \pm\infty$, $x \rightarrow \infty$), the tangent becomes $V_\gamma \sim \partdif{}{t}$; i.e., matter loses its kinetic energy and comes to rest.
Therefore, at late times, the light ray is not caught any more as the light moves outwards faster than the matter (see also Fig.~\ref{fig:causalstructure}).
We can do form a schematic plot of the light rays where ingoing and outgoing ones have $45^\circ$ angles w.r.t. the horizontal and vertical axes.
Most importantly, they are orthogonal to each other in this depiction.
Usually, this is automatically done by finding Kruskal-Szekeres-like coordinates and performing the causal compactification.
Due to the Planck-scale breaking of the diffeomorphism invariance, this is not possible here, and we can only sketch this.
The result is given in Fig.~\ref{fig:penrose}.
All the features worked out in the previous subsections are nicely visible.
Ingoing light rays originate at infinity and move towards $x_{min}$.
Outgoing light rays, which lie outside of $x_{out}$ originate in principle at $x_{out}$, which is not the case here as they first intersect the curve $L(t)$.
In their future, they escape to infinity or intersect $L(t)$.
More interesting is the region $x_{in} < x < x_{out}$.
As was shown in Sec.~\ref{sec:radialgeos}, the outgoing light rays all tend (in their future and past) to $x_{in}$.
Also, this is faithfully represented by the schematic Penrose diagram.
Nevertheless, they never reach $x_{in}$, as they first intersect the curve $L(t)$.
The case is similar for light rays with $x_{min} < x < x_{in}$, which ignoring the intersection with the matter, originate at $x_{min}$ and terminate at $x_{in}$.
This gives a precise picture of the causal structure of the model \cite{KellyBlackholecollapse}.

\section{Resolving the Spacetime Discontinuity}\label{sec:discont}

It is possible to avoid the super-luminal speed of the matter and the shock wave completely.
This can be done by removing the discontinuity in the metric.
The source of the issue lies in the choice $E^a = x^2$ in finding the true Hamiltonian.
This is an excellent gauge in classical general relativistic systems, but it is tricky in bouncing LQC-like models.
Comparing to other models as \cite{VakiliClassicalpolymerizationof,CorichiLoopquantizationof,ModestoSemiclassicalLoopQuantum,BoehmerLoopquantumdynamics,BenAchourPolymerSchwarzschildBlack,AshtekarQuantumExtensionof,AshtekarQuantumTransfigurarationof,BodendorferEffectiveQuantumExtended,BodendorferMassandHorizon,Bodendorferbvtypevariables,AssanioussiPerspectivesonthe,KellyEffectiveloopquantumgravity,KellyBlackholecollapse,GambiniSphericallysymmetricloop,GeillerSymmetriesofthe,SartiniQuantumdynamicsof,GanPropertiesofthe}, one generically finds that $E^a(\lambda)$ is a non-invertible function with two branches.
Here $\lambda$ is some parameter choice that appears if there the Hamiltonian is not completely gauge fixed.
De-parametrising the solutions includes the inversion $\lambda(E^a)$, which generically has two branches, the one before and the other after the bounce.
Therefore, when choosing the gauge $E^a = x^2$ one has to track very carefully all appearing signs as usually, one sign choice is the pre- and the other the after-bounce spacetime branch.

Let us enter the details of the model to point out where this sign might help to solve the spacetime discontinuity.
Recall that the equations of motion (cfr. Eqs.~\eqref{eq:b} and \eqref{eq:rho}) are given by
\begin{align}
	\dot{b} =& -\frac{1}{2\gamma \Delta x} \partdif{}{x} \left(x^3 \sin^2\left(\frac{\sqrt{\Delta} b}{x}\right)\right)\;,
	\label{eq:b2}
	\\
	\rho =& \frac{1}{8 \pi G \gamma^2 \Delta x^2} \partdif{}{x} \left(x^3 \sin^2\left(\frac{\sqrt{\Delta} b}{x}\right)\right) \;,\label{eq:rho2}
\end{align}
\noindent
where $\partial_x \rho = 0$ for $x \neq L(t)$ in the OS case (edge effects are again neglected).
From energy conservation, it follows that $M = 4\pi \rho L^3/3 =const.$ (see \cite{KellyBlackholecollapse}), which relates $\rho$ with the radius of the outermost shell $L$.
Eq.~\eqref{eq:rho2} can easily be integrated leading to
\begin{equation}\label{eq:solb}
	\sin^2\left(\frac{\sqrt{\Delta} b}{x}\right) = \begin{cases}
		\frac{\rho(t)}{\rho_c} + \frac{D}{x^3} &,\; \rho \neq 0 
		\\
		\frac{C}{x^3} &,\; \rho = 0
	\end{cases}
\;.
\end{equation}
\noindent
Here $C$ and $D$ are integration constants.
They can be fixed by suitable matching conditions.
Usually one chooses $D = 0$ as this contribution diverges at $x = 0$.
As $x \ge x_{min} > 0$, this argument does not apply here.
Still, we choose here $D = 0$, as this does not affect the discussion and results.
We also define $\rho_c = 3/(8 \pi G \gamma^2 \Delta)$.
The solution in the region $\rho = 0$ corresponds to the vacuum metric, which is the same as the one found in \cite{KellyEffectiveloopquantumgravity}.
However, in the original work \cite{KellyBlackholecollapse}, the square root was taken and a negative sign was fixed.
We try to avoid this as long as possible and see how this can be used to resolve the discontinuity.
Taking the time-derivative of Eq.~\eqref{eq:solb} and inserting Eq.~\eqref{eq:b2} leads (in fact, even for $D \neq 0$) to
\begin{align}
	\frac{\dot{\rho}}{\rho} =& -\frac{3}{\gamma \sqrt{\Delta}} \sin\left(\frac{\sqrt{\Delta} b}{x}\right) \cos\left(\frac{\sqrt{\Delta} b}{x}\right) = 3 \frac{N_{mat}^x}{x} \;,
	\\
	\frac{\dot{L}}{L} =& \frac{1}{\gamma \sqrt{\Delta}} \sin\left(\frac{\sqrt{\Delta} b}{x}\right) \cos\left(\frac{\sqrt{\Delta} b}{x}\right) = -\frac{N_{mat}^x}{x} \;,
	\label{eq:Nxmatter}
\end{align}
\noindent
where Eq.~\eqref{eq:Nx} is used for the last equality.
This can be viewed as the solution for the spacetime coefficient $N^x_{mat}$ and thus there is no need to choose a sign here.
It is obviously needed once the equation for $L$ has to be solved, but not to determine the spacetime metric in the matter region.
The equation for $L$ in a self-contained\footnote{One might wonder how this equation matches Eq.~\eqref{eq:Nxmatter}, which depends on $x$ explicitly, while Eq.~\eqref{eq:dotL} does not. Comparing with the solution Eq.~\eqref{eq:solb} (for $D = 0$) the expression $N_{mat}^x/x$ is actually independent of $x$. Therefore it is possible to simply insert any value for $x$, most suitably $x = L(t)$.} form becomes
\begin{equation}\label{eq:dotL}
	\left(\frac{\dot{L}}{L}\right)^2 = \frac{8 \pi G}{3} \rho \left(1- \frac{\rho}{\rho_c}\right) \;,
\end{equation}
with $\rho = 3M/(4\pi L^3)$.
As this expression comes in a quadratic form, there is also no explicit choice of sign here.

The situation is more subtle in the vacuum region i.e., $\rho = 0$ as there are also two sign options.
The solution boils down to the solutions of \cite{KellyEffectiveloopquantumgravity} simply with the gauge choice $N^2 = 1$ which is implied by the choice of the matter field as time.
Nevertheless, it is instructive to solve the dynamical equations of \cite{KellyEffectiveloopquantumgravity} in the most general sense as this allows getting an overview of all possible solutions.
This analysis is shifted to the appendix~\ref{app:fullsolutions} and coincides with the following steps.
The sign we have to choose here corresponds to the sign that has to be chosen in the equations $N^2 = 1$.
First, we need to fix a sign of $\sin\left(\frac{\sqrt{\Delta} b}{x}\right)$, which we call $s = 1$ for $\frac{\sqrt{\Delta} b}{x} \in \left[0, \pi\right]$ and $s = -1$ for the interval $\left[\pi,2\pi\right]$.
Note that $\frac{C}{x^3} \in \left(0,\infty\right)$ (for $C >0$).
We can then fix $x \in \left[x_{min},\infty\right)$ and therefore $C = x_{\min}^3$ to keep this expression\footnote{Note that this does not fix the integration constant $C$ yet, as up to here $x_{min}$ is just another name and its relation to $M$ is not determined.} bounded between $0$ and $1$.
Therefore, this restricts $\sin\left(\frac{\sqrt{\Delta} b}{x}\right)$ to only one half of the respective branch i.e., $\frac{\sqrt{\Delta} b}{x}\in \left[0,\pi/2\right]$ or $\frac{\sqrt{\Delta} b}{x}\in \left[\pi/2,\pi\right]$ but the branch is never crossed.
Respectively, for the branch $\left[\pi,2\pi\right]$ with the negative sign, the solution remains on a quarter of the circle, and we can safely write
\begin{equation}
	\cos\left(\frac{\sqrt{\Delta} b}{x}\right) = c \sqrt{1-\sin^2 \left(\frac{\sqrt{\Delta} b}{x}\right)} \;,
\end{equation}
\noindent
where $c = \text{sign}(\cos)$ and this sign remains fixed during the entire evolution\footnote{In writing $\cos(x) = \pm\sqrt{1-\sin(x)^2}$ there is also a sign choice needed. This sign remains fixed during the evolution, which simplifies the analysis.}. 
For the vacuum region it is then
\begin{equation}\label{eq:Nxvacsign}
	N_{vac}^x=-\frac{x}{\gamma \sqrt{\Delta}} \sin\left(\frac{\sqrt{\Delta} b}{x}\right) \cos\left(\frac{\sqrt{\Delta} b}{x}\right) = -s\cdot c\,\sqrt{\frac{x_{min}^3}{\gamma^2 \Delta x}\left(1-\frac{x_{min}^3}{x^3}\right)} \;,
\end{equation}
which obviously has the same sign for all $t$.
Therefore, while the matter region does change the sign of $N^x_{mat}$, this is a priori not possible for the vacuum region.
However, it is possible to freely choose $\text{sign}\left(N_{vac}^x\right) = \pm 1$.

We have to match the matter and vacuum regions at the matter surface $x = L(t)$.
For the matter region, this gives
\begin{equation}
\left.\dd s_{mat}^2\right|_{x = L(t)} = -\left(1-2(N_{mat}^x)^2-2 \dot{L}N_{mat}^x\right) \dd t^2 + L^2(t) \dd \Omega^2 = -\dd t^2 + L^2(t) \dd \Omega^2 \;,
\end{equation}
\noindent
where Eq.~\eqref{eq:Nxmatter} and $N_{mat}^x(L) = -\dot{L}$ were used.
Instead for the vacuum region, this gives
\begin{equation}
	\left.\dd s_{vac}^2\right|_{x = L(t)} = -\left(1-2(N_{vac}^x)^2-2 \dot{L}N_{vac}^x\right) \dd t^2 + L^2(t) \dd \Omega^2 \;.
\end{equation}
\noindent
It is now necessary to analyse $\dot{L}$ and Eq.~\eqref{eq:dotL} in more detail.
Eq.~\eqref{eq:dotL} can be re-written as
\begin{equation}
	|\dot{L}| = \sqrt{\frac{R_s}{L}\left(1-\frac{R_s \Delta \gamma^2}{L^3}\right)} \;,
\end{equation}
\noindent
with $R_s = 2 G M$.

Demanding continuity across the surface $x = L(t)$ gives
\begin{align}
	\left.\dd s_{mat}^2\right|_{x = L(t)} =& \left.\dd s_{vac}^2\right|_{x = L(t)} 
	\notag
	\\
	\Rightarrow 0 =& \left(N^x_{vac}\right)^2 + \dot{L} N^x_{vac} \;.
\end{align}
\noindent
This can only be solved for 
\begin{align}
	\text{sign}\left(\dot{L}\right) &= -\text{sign}\left(N_{vac}^x\right)
	\\
	x_{min}^3 &= R_s \gamma^2 \Delta \quad \Rightarrow \quad \dot{L} = -N_{vac}^x(L) \;,
\end{align}
which fixes all relevant sign choices and the integration constant $C = x_{min}^3$ in terms of the mass $M$.

The crucial point is then that the sign of $N_{vac}^x$ has to be different in the collapsing phase and the re-expanding one.
As we argued before the sign of $N_{vac}^x$ remains fixed during the whole evolution.
This seems to be a contradiction and might lead to the solution of accepting a discontinuous shock wave as discussed in \cite{KellyBlackholecollapse}.
There exists another solution avoiding any discontinuity by extending the spacetime across $x_{min}$ by a bounce.

We assume first that $\text{sign}\left(N_{vac}^x\right) = 1$, as was chosen before and in \cite{KellyBlackholecollapse,KellyEffectiveloopquantumgravity}.
This gives the metric in ingoing Painlevé-Gullstrand coordinates.
The metric is continuous across the surface $x = L(t)$ in the collapsing phase.
To make it also continuous in the re-expanding phase, we need the exact same metric but with $\text{sign}\left(N_{vac}^x\right) = -1$.
This is nothing other than the time-reversed metric i.e., outgoing Painlevé-Gullstrand coordinates.
We can now simply identify the two spacetimes regions with $\text{sign}\left(N_{vac}^x\right) = \pm1$ across $x = x_{min}$, where the sign of $N_{mat}^x$ changes.
This identification is then continuous as 
\begin{align}
	\left.\dd s_{vac,-}^2\right|_{x=x_{min}} =& \left.\dd s_{vac,+}^2\right|_{x=x_{min}}
	\notag
	\\
	\Leftrightarrow \left(N_{vac,-}^x\right)^2 =& \left(N_{vac,-}^x\right)^2 \;,
\end{align}
\noindent
which is satisfied as $N_{vac,\pm}^x$ is squared.
The spacetime is even at least once continuously differentiable at $x = x_{min}$ as the extrinsic curvature
\begin{equation}
K_{x = const.} = \sqrt{1-\left(N^x_{vac}\right)^2} N_{vac}^x \frac{\dd N_{vac}^x}{\dd x} \dd t^2 + \sqrt{1-\left(N^x_{vac}\right)^2} x \dd\Omega^2 \;,
\end{equation}
\noindent
coincides for both signs of $N^x_{vac}$ at $x= x_{min}$.
Inserting $x = x_{min}$ yields in both cases 
\begin{equation}
K_{x = x_{min}}^\pm = -\frac{3R_s}{2 x_{min}^2} \dd t^2 + x_{min} \dd \Omega^2 \;.
\end{equation}
Therefore this identification at $x=x_{min}$ is valid from a continuity point of view.
The re-expanding branch of the matter then takes place in the time-reversed region with $\text{sign}\left(N_{vac}^x\right) = -1$, which gives a continuous bounce everywhere.
Following, therefore, the collapsing matter, $N_{vac}^x$ indeed changes its sign.
The metric is then continuous, and $b$ is also continuous.
According to the arguments given in \cite{KellyBlackholecollapse}, this shows further that the neglect of edge effects is a valid approximation here.
We can use a new coordinate $\lambda$ defined by $x^2 = \left(\lambda^2/2 + x_{min}\right)^2$, which allows us to describe the full spacetime at once.
The line element of the vacuum metric then reads
\begin{equation}\label{eq:metriclambda}
	\dd s^2 = -\left(1-\left(N_{vac}^x(\lambda)\right)^2\right) \dd t^2 +2 N_{vac}^x(\lambda)\left|\lambda\right| \dd t \dd \lambda + \lambda^2 \dd \lambda^2 + \left(\frac{\lambda^2}{2}+x_{min}\right)^2 \dd \Omega^2 \;,
\end{equation}
\noindent
with
$$
N_{vac}^x = \sqrt{\frac{R_s}{\frac{\lambda^2}{2} + x_{min}} \left(1-\frac{x_{min}^3}{\left(\frac{\lambda^2}{2}+x_{min}\right)^3}\right)} \,.
$$
\noindent
With $\lambda \in \mathbb{R}$, we find indeed that $g_{tx} > 0$ for $\lambda > 0$ and $g_{t x} < 0$ for $\lambda < 0$.
This is possible, as the relation of $x$ and $\lambda$ is non-invertible, and on one branch one has to choose a different sign respective the other sign respectively the other\footnote{Note the formal similarity to the areal radius in \cite{DAmbrosioHowinformationcrosses}.}.
This again gives rise to a bounce at a minimal radius and makes connection to almost all previous LQG-inspired black hole models \cite{VakiliClassicalpolymerizationof,CorichiLoopquantizationof,ModestoSemiclassicalLoopQuantum,BoehmerLoopquantumdynamics,BenAchourPolymerSchwarzschildBlack,AshtekarQuantumExtensionof,AshtekarQuantumTransfigurarationof,BodendorferEffectiveQuantumExtended,BodendorferMassandHorizon,Bodendorferbvtypevariables,AssanioussiPerspectivesonthe,KellyEffectiveloopquantumgravity,KellyBlackholecollapse,GambiniSphericallysymmetricloop,GeillerSymmetriesofthe,SartiniQuantumdynamicsof,GanPropertiesofthe}.

Another cross-check is the continuity along $t = const.$ surfaces in the vacuum region i.e., $x > L(t)$, and in fact $t > 0$.
The induced metric is then simply
\begin{equation}
	\left.\dd s^2\right|_{t=const.} = \dd x^2 + x^2 \dd\Omega^2 \;,
\end{equation}
\noindent
which is obviously continuous, even across $x = x_{min}$ as it is independent of $t$ and also $x$.
More tricky is the extrinsic curvature, which yields
\begin{equation}
	\left.K\right|_{t=const.} = \pm \frac{\dd N_{vac}^x}{\dd x} \dd x^2 \pm x N_{vac}^x \dd\Omega^2 \,.
\end{equation}
\noindent
The angular part is continuous across $x_{min}$ as $N_{vac}^x(x_{min}) = 0$, while the radial part diverges (cfr. Eq.~\eqref{eq:psiprime}).
This can simply be interpreted as a bug of the chart $x$ which ends at $x_{min}$.
Changing instead to the coordinate $\lambda$, which can be extended across $x_{min}$, we find
\begin{equation}
	\left.K_{\lambda \lambda} \right|_{t=const.} = |\lambda| \frac{\dd N_{vac}^x}{\dd \lambda} \xrightarrow{\lambda \rightarrow 0} 0 \;,
\end{equation}
\noindent
which is straightforward to prove to be continuous across the bounce.

\begin{figure}[h!]
	\centering\includegraphics[width=9cm]{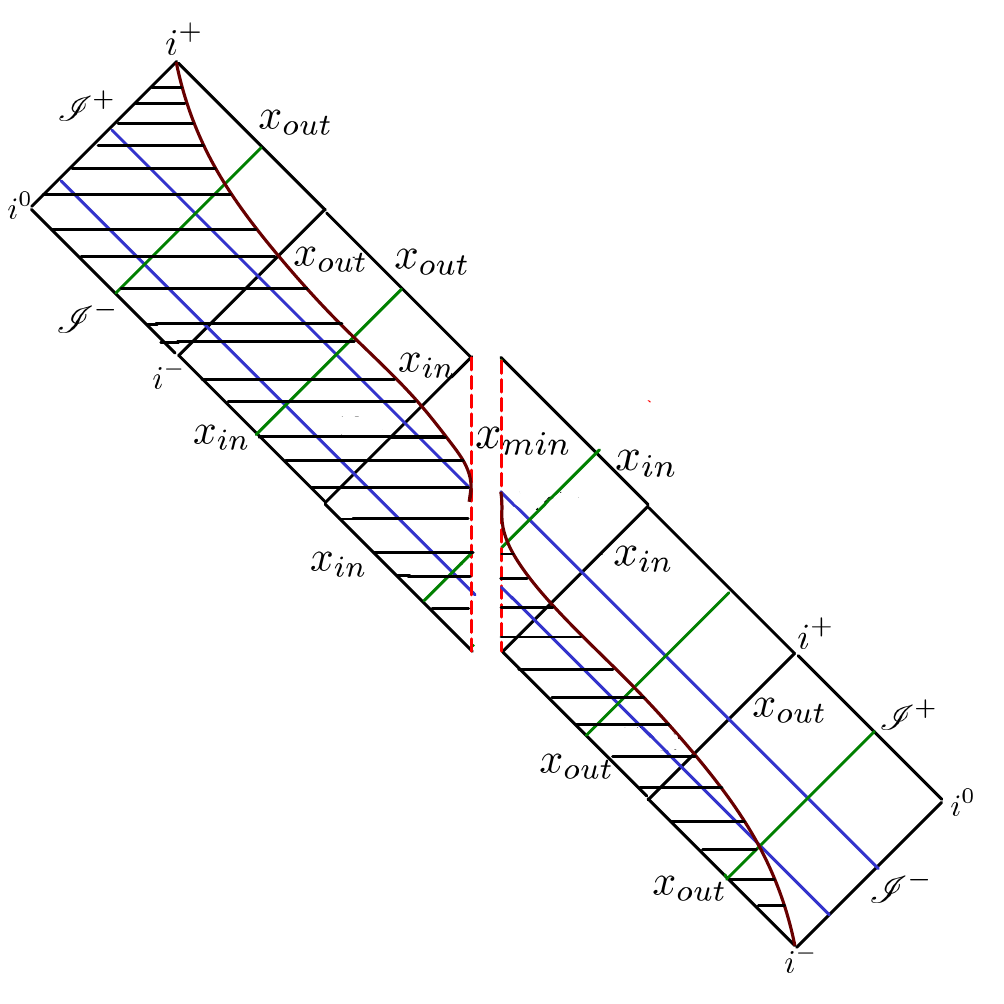}
	\caption{Schematic Penrose diagram of the continuous collapse process. The two different vacuum solutions are identified at $x = x_{min}$. The matter can travel across this surface and remains in time-like motion throughout the full process. The region where the matter solution is relevant is shaded out. The solution is now incomplete as outgoing light rays can e.g. exit via $x_{in}$. Some sample ingoing (blue) and outgoing (green) light rays are drawn. The $\lambda$-chart and the metric Eq.~\eqref{eq:metriclambda} covers the full (vacuum region) diagram.}
	\label{fig:penrose2}
\end{figure}

This identification has a tremendous effect on the global causal structure.
First of all, we find that 
\begin{equation}
	g\left(V_\gamma,V_\gamma\right) = -\left(1-N^x_{vac}(L(t))^2\right) + 2 N^x_{vac}(L(t)) \dot{L}(t) + \dot{L}(t)^2 = -1 \,, 
\end{equation}
\noindent
and therefore the evolution of matter is time-like during the full evolution, even in the re-expanding phase.
As the spacetime region with $\text{sign}\left(N_{vac}^x\right) = -1$ is just the time-reverse of the other sign, all previous analysis of light rays remains unaltered, expect that the ingoing and outgoing light rays are exchanged.
It is therefore again possible to sketch a Penrose diagram, which is now given in Fig.~\ref{fig:penrose2}.
The minimal-radius surface $x = x_{min}$ now plays the role of a transition surface.
However, this surface is time-like, rather than space-like and also does not transition between a trapped and an anti-trapped region.
The areal radius reaches a minimal value and then continues into a parallel universe, the second branch of which is usually found in LQC-like black hole models.
Note that the spacetime is now incomplete.
Before all light rays with $x < x_{out}$ were caught again by the matter surface (cfr. Fig.\ref{fig:penrose}).
This is not true any more; outgoing light rays reach $x_{in}$ as is visible from the sketched Penrose diagram in Fig.~\ref{fig:penrose2} and the computations in Sec.~\ref{sec:radialgeos}.
The inner horizon is no longer covered by the surface $x = L(t)$ as in the original description.

\begin{figure}[h!]
	\centering\includegraphics[width=15cm]{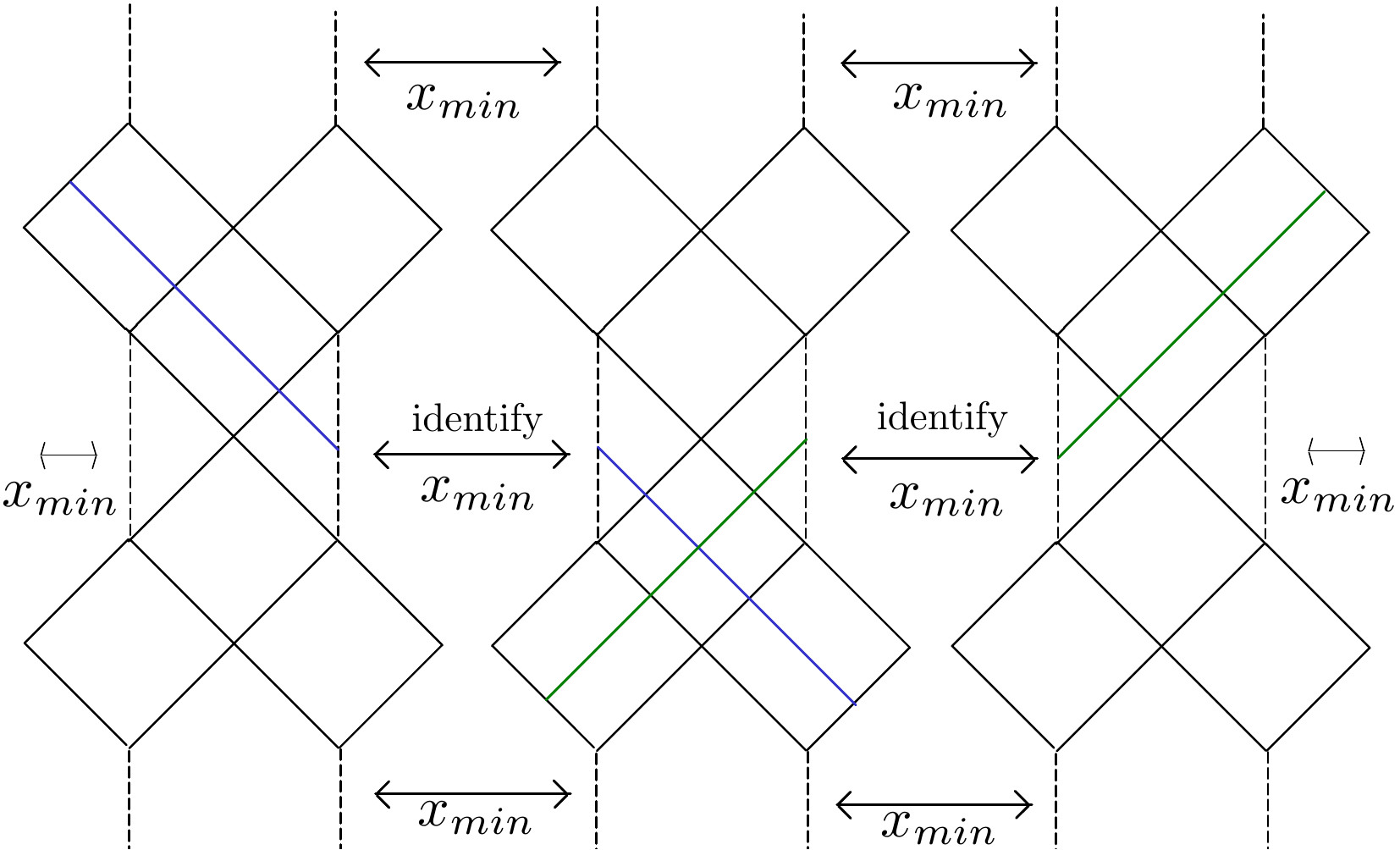}
	\caption{Schematic Penrose diagram of the vacuum solution without matter. This can be achieved by identifying shared regions of the spacetime solutions with $\text{sign}\left(N^x\right) = \pm1$ and additionally identifying them across $x=x_{min}$. The global structure is infinitely extended and similar to a Reissner-Nordström black hole but also infinitely extended in the vertical direction.}
	\label{fig:penrose3}
\end{figure}

One is tempted to extend the coordinates (the possibility of which is not clear due to non-covariance) to ``global Kruskal-Szekeres'' coordinates and cover also the regions behind $x_{in}$.
This is not needed as we can match several patches together and identify them appropriately.
We find the two solutions in ingoing and outgoing Painlevé-Gullstrand coordinates and can now repeatedly cover shared regions of them.
This can be made mathematically exact.
Consider the two solutions
\begin{subequations}
	\begin{align}
		\dd s^2_+ =&-\left(1-\left(N_{vac}^x\right)^2\right) \dd t_+^2 + 2 N_{vac}^x \dd t_+ \dd x + \dd x + x^2 \dd \Omega^2 \;,
		\\
		\dd s^2_- =&-\left(1-\left(N_{vac}^x\right)^2\right) \dd t_-^2 - 2 N_{vac}^x \dd t_- \dd x + \dd x + x^2 \dd \Omega^2 \;.
	\end{align}
\end{subequations}
\noindent
It is possible to identify the regions $x > x_{out}$ by simply mapping $x \mapsto x$ and relating the times according to
\begin{equation}
	t_+ = t_- + 2 \int_{x_{ref}}^x \frac{N_{vac}^x}{1-\left(N_{vac}^x\right)^2} \dd x' \;.
\end{equation}
\noindent
This is a coordinate transformation, but it leaves the spacetime scalars unaltered as the lapse function is only mapped from $N = 1\mapsto N = -1$ (see App.~\ref{app:fullsolutions} for details. 
$N^x \propto N$, and the sign change in $N$ changes the sign of $N^x$.)
This is therefore effectively a time-reverse in a static spacetime.
Note that this transformation diverges if $x_{ref}$ and $x$ lie on different sides of the horizon, which is expected, as the two charts only coincide for the regions $x > x_{out}$ or $x_{in} < x < x_{out}$ or $x_{min} < x < x_{in}$.
Regardless, this is enough to glue sufficiently many of these patches together to extend all geodesics and gain a global extension of the spacetime.
This procedure is exact and always possible, though it carries some ambiguity regarding the chosen topology of spacetime. 
This allows one to construct the full vacuum solution given by the equations of \cite{KellyEffectiveloopquantumgravity} (without matter) and gives an extension of the sketched Penrose diagram, which has now a Reissner-Nordström-like structure, but is infinitely extended across all $x= x_{min}$ surfaces (see Fig.~\ref{fig:penrose3}).
The Penrose diagram would therefore be infinitely extended in both time and space directions.
The overall picture of the collapse is then very similar to what was found in \cite{MuenchEffectiveQuantumDust}.
The cost of this continuous spacetime metric is the fact that the Penrose diagram is, even including matter, not cut down to a finite diagram.
This gives the hint that important global effects are missing, e.g. Hawking evaporation, which potentially merges all vacuum regions after the black hole evaporation \cite{AshtekarBlackHoleevaporation,HaggardBlackHoleFireworks,BianchiWhiteHolesasRemnants,ChristodoulouPlanckstartunneling,ChristodoulouCharacteristicTimeScales,Martin-DussaudEvaporationgblackto,DAmbrosioTheEndof}.
                                                               
\section{Conclusions}\label{sec:conclusions}

In this paper the causal structure of the LQG OS collapse model \cite{KellyBlackholecollapse}, related to \cite{KellyEffectiveloopquantumgravity}, has been analysed.
A speciality of this model is that the spacetime is not diffeomorphism invariant at the Planck regime; it depends on the choice of lapse function, although it is suppressed by factors of the area gap $\Delta$.
Then, as the metric is not independent of the choice of coordinates, it is not straightforwardly possible to construct global conformally compactified coordinates and derive a Penrose diagram.
Nevertheless, it is possible to study light-like radial geodesics, which capture the main causal features.
Note that this strategy could also be applied to other models, which break diffeomorphism invariance at the quantum level as in \cite{GambiniSphericallysymmetricloop,GambiniLQBHextensions}.
This was first done for the vacuum region of the spacetime, which coincides with the eternal description of \cite{KellyEffectiveloopquantumgravity}.
It turns out that there are two apparent horizons $x_{in}$ and $x_{out}$, which in the eternal case are Killing and causal horizons.
As in the collapse model \cite{KellyBlackholecollapse}, spacetime is dynamical due to the presence of matter, the presence of this causal horizon has to be rediscussed.
It was found that it is not present in the collapsing case.
Radial ingoing light rays move inwards and pass both horizons until they hit the minimal radius surface $x_{min}$, where the vacuum spacetime ends.
This is no problem as the light rays are captured by the collapsing matter surface before reaching this point and then propagate regularly in the matter region of spacetime.
Outgoing light rays emitted outside of $x_{out}$ reach out to infinity and in their past originate at the outer horizon $x_{out}$, both within infinite coordinate time.
Instead, light rays starting between the inner and outer horizons $x_{in} < x < x_{out}$ arrive in the future always at $x_{in}$ and in their past at $x_{out}$, also in infinite coordinate time.
Again, before reaching these horizons they are captured by the matter surface and continue their propagation according to the matter region metric.
The last case are outgoing light rays starting at $x_{min} < x < x_{in}$, which in the past originate at $x_{min}$ and evolve towards $x_{in}$ in infinite time.
The fact that these horizons are reached only within infinite coordinate time indicates that the chart does not cover regions beyond these points.
This can be understood as interpreting the eternal metric as a part of a Reissner-Nordstöm-like causal structure.

In the next step, the matter trajectory was analysed.
There is a conceptional miss-match since $x_{out}$ is a causal horizon for the eternal black hole but not for the collapsing spacetime.
It was shown that it is possible that light rays can enter the matter region at $x<x_{out}$ expand together with the matter and exit at $x>x_{out}$, thus causally connecting the black hole interior and exterior.
The causal horizon disappears in the collapsing model.
The question remains of how this can happen, and it was worked out that the matter is re-expanding faster than light out of the black hole interior.
The norm of its tangent becomes space-like for a relevant part of the evolution and thus can move out of the outer horizon.
This happens at the re-expanding branch, where a shock wave \cite{KellyBlackholecollapse} appears and the spacetime metric is discontinuous across the matter surface.
Note that this is the perception of a vacuum-region observer only.
From an observer inside the collapsing matter, the surface would travel along a time-like path, which gives no contradictions due to the discontinuity of the spacetime across the matter surface in the re-expanding branch.
This is consistent with earlier work \cite{BenAchourBouncingcompactobjectsI,BenAchourBouncingcompactobjectsII,BenAchourConsistentblackto,MuenchEffectiveQuantumDust,SchmitzExteriorstobouncing}, which all conclude that a bounce has to happen outside of a causal horizon of the eternal metric or it will leave in a parallel universe if one insists on a continuous and time-like collapse.
Both of these assumptions are violated in \cite{KellyBlackholecollapse}, which gives rise to this shock wave solution.

Finally, it was shown that the dynamical equations of \cite{KellyBlackholecollapse,KellyEffectiveloopquantumgravity} also allow continuous solutions, where the matter does follow a time-like trajectory everywhere.
Therefore it was shown that the time-reversed vacuum metric is also a solution of these equations and a once continuously differentiable matching between these two vacuum metrics at the minimal radius surface $x = x_{min}$ is possible.
The collapsing matter can therefore continuously follow its collapse across the surface $x=x_{min}$ and re-expand in the parallel universe.
The process is then continuous everywhere and justifies the neglect of edge effects.
However, the cost of this is an enormously more complicated causal structure.
The collapsing matter surface is not cutting out additional regions and light and matter can e.g. escape across $x_{in}$ and the parallel universe.
This picture is again very familiar with other LQG-inspired black hole modes \cite{VakiliClassicalpolymerizationof,CorichiLoopquantizationof,ModestoSemiclassicalLoopQuantum,BoehmerLoopquantumdynamics,BenAchourPolymerSchwarzschildBlack,AshtekarQuantumExtensionof,AshtekarQuantumTransfigurarationof,BodendorferEffectiveQuantumExtended,BodendorferMassandHorizon,Bodendorferbvtypevariables,AssanioussiPerspectivesonthe,KellyEffectiveloopquantumgravity,KellyBlackholecollapse,GambiniSphericallysymmetricloop,GeillerSymmetriesofthe,SartiniQuantumdynamicsof,SchmitzExteriorstobouncing,GanPropertiesofthe} and the collapse model \cite{MuenchEffectiveQuantumDust} where an infinite tower of repeating Penrose diagrams appears and the matter does not cut out all irrelevant regions.
This allows the speculation that there is something very essential missing to determine a physically reasonable black hole spacetime, which most likely is black hole evaporation \cite{AshtekarBlackHoleevaporation,HaggardBlackHoleFireworks,BianchiWhiteHolesasRemnants,ChristodoulouPlanckstartunneling,ChristodoulouCharacteristicTimeScales,Martin-DussaudEvaporationgblackto,DAmbrosioTheEndof}.

Further, having all these patches of spacetimes in ingoing and outgoing Painlevé-Gullstrand coordinates allows for the complete eternal and vacuum metric.
The resulting spacetime has a causal structure that is similar to a Reissner-Nordstöm black hole,
although the singularity is resolved and instead replaced by $x_{min}$ where a minimal radius is reached and the spacetime is regular \cite{KellyEffectiveloopquantumgravity}.
The different Reissner-Nordström patches can be identified at $x = x_{min}$ leading to an infinite extended causal structure in the time and space directions.
All this is supplemented by the computations in App.~\ref{app:fullsolutions}, where the eternal equations of \cite{KellyEffectiveloopquantumgravity} were solved for any gauge choice of $N$, which shows that the time-reversed metric is a valid solution.
It also allows us to make the breaking of the diffeomorphism invariance explicit, as spacetime scalars, (such as e.g. $R_{\mu \nu \alpha \beta} R^{\mu \nu \alpha \beta}$) depend now on the lapse function $N$, although the dependence is suppressed by $\Delta$.
It should be emphasised that this seems to be a general issue in LQG-related black hole models \cite{BojowaldAno-goresult}.
A first attempt to understand this in terms of quantum discreteness was discussed in \cite{GambiniLQBHextensions}.
This also allows to find different coordinate solutions discussed in \cite{KellyEffectiveloopquantumgravity} and \cite{GambiniSphericallysymmetricloop} as special cases of this general solution.
Further it is possible to make contact with \cite{BodendorferMassandHorizon} and analyse the number of physically relevant integration constants.
Consistent with the observations in \cite{BodendorferMassandHorizon} there exists only one physical integration constant, related to the mass, as there are no fiducial cell depended polymerisation scales\footnote{Although this is not a good notion in this setting, which is free of fiducial cell issues, rephrased in the language of the static spacetimes of \cite{BodendorferMassandHorizon} this would correspond to a fiducial cell independent polymerisation --- if it is possible to relate these.}.
There is a second integration constant that can always be absorbed in the lapse $N$ and does not appear in spacetime scalars.
This allows for speculating that $N$ itself might be a full function of ``initial conditions'', which might be fixed due to reasonable physical conditions and avoids then the breaking of diffeomorphism invariance due to the choice of physically preferred situations.
This certainly should be understood in future work.

In general, we saw that the model \cite{KellyBlackholecollapse} allows two pictures of black hole collapse.
On the one hand, there is the shock wave solution, which has the drawback of spacetime discontinuity and space-like matter evolution.
Nevertheless, it has promising features, such as a nice compact causal structure without any causal horizon.
Besides this, it is then possible to compute the lifetime of a black hole without the need of including black hole evaporation, which is proportional to $M^2$ \cite{KellyBlackholecollapse} and makes contact with \cite{ChristodoulouPlanckstartunneling}, which is significantly smaller than the evaporation time scale $M^3$ and the Page time \cite{PageAverageentropyof}.
One might therefore argue that Hawking evaporation effects are sub-dominant and negligible.
Due to the avoidance of Cauchy horizons, which are all cut out by the matter, stability issues as discussed in \cite{Carballo-RubioOntheviability} do not apply. 
These features were not achieved in any LQG black hole model so far, at least when Hawking evaporation is neglected.
Therefore, it still might be possible to insist on the discontinuous metric and the shock wave.
However, it would require us to physically justify this discontinuity and the super-liminal re-expansion from the vacuum-region perspective.

On the other hand, there is the continuous solution.
This solution has the advantage of being continuous and admitting a time-like matter evolution everywhere.
Nevertheless, this is at the cost of an infinite extended causal structure, but also makes nice contact with previous bouncing black hole models, where two branches appear \cite{VakiliClassicalpolymerizationof,CorichiLoopquantizationof,ModestoSemiclassicalLoopQuantum,BoehmerLoopquantumdynamics,BenAchourPolymerSchwarzschildBlack,AshtekarQuantumExtensionof,AshtekarQuantumTransfigurarationof,BodendorferEffectiveQuantumExtended,BodendorferMassandHorizon,Bodendorferbvtypevariables,AssanioussiPerspectivesonthe,KellyEffectiveloopquantumgravity,KellyBlackholecollapse,GambiniSphericallysymmetricloop,GeillerSymmetriesofthe,SartiniQuantumdynamicsof,GanPropertiesofthe}.
The main difference between these approaches is that the transition surface is space-like and does not transition a trapped region into an anti-trapped region. 
It remains to say that in this case there is still hope that Hawking radiation will allow us to avoid these infinite exteriors as discussed by \cite{AshtekarBlackHoleevaporation,HaggardBlackHoleFireworks,BianchiWhiteHolesasRemnants,ChristodoulouPlanckstartunneling,ChristodoulouCharacteristicTimeScales,Martin-DussaudEvaporationgblackto,DAmbrosioTheEndof}.
Black hole evaporation is therefore a necessary ingredient to gain a full understanding of a black hole life cycle.

The physical question remains: Does the matter bounce out of a black hole as a shock wave with super-luminal speed or does it not bounce out at the cost of an infinite tower of exterior regions?
For this, the physical consequences should be worked out in the future and the breaking of diffeomorphism invariance should be understood better.

\section*{Acknowledgements}

JM gratefully acknowledges insightful discussions with Edward Wilson-Ewing, Jarod G. Kelly, and Robert Santacruz and their valuable feedback on this work.
This publication was made possible through the support of the ID\# 61466 grant from the John Templeton Foundation, as part of the “The Quantum Information Structure of Spacetime (QISS)” Project (qiss.fr). The opinions expressed in this publication are those of the author(s) and do not necessarily reflect the views of the John Templeton Foundation.

\appendix
\section{Full Space of Vacuum Solutions}\label{app:fullsolutions}

We come back to the more general setting of \cite{KellyEffectiveloopquantumgravity}, where no matter is included.
Consequently, there is no natural time gauge fixing coming from the matter field.
This makes the equations of motion more general as the lapse $N$ is not fixed.
There is still the gauge $E^a = x^2$ applied, and the quantisation scheme is exactly the one described in Sec.~\ref{sec:model}.
The spacetime metric still has the form of Eq.~\eqref{eq:metric}, but the expression in Eq.~\eqref{eq:Nx} for the shift $N^x$ has an additional factor of $N$ i.e., 
\begin{equation}\label{eq:Nxfull}
	N^x = -\frac{N x}{\gamma \sqrt{\Delta}} \sin\left(\frac{\sqrt{\Delta} b}{x}\right)\cos\left(\frac{\sqrt{\Delta} b}{x}\right) \;.
\end{equation}
\noindent
Spacetime dynamics is generated by $H = \int\dd x N\H^{eff}$, where $\H^{eff}$ is given by Eq.~\eqref{eq:Heff}, which now is not a true Hamiltonian, but rather a constraint $\H^{eff} \approx 0$.
The equations of motion Eq.~\eqref{eq:EoM} are modified by a factor $N$ and thus read
\begin{subequations}\label{eq:EoMfull}
	\begin{align}
		\dot{E}^b =& - \frac{x^2}{2 \gamma \sqrt{\Delta}} \partial_x \left(\frac{N E^b}{x}\right) \sin\left(\frac{\sqrt{\Delta} b}{x}\right)\cos\left(\frac{\sqrt{\Delta} b}{x}\right) = \frac{x}{2 N} N^x \partial_x \left(\frac{N E^b}{x}\right)\label{eq:dotEbfull} \;,
		\\
		\dot{b} =& \, \frac{\gamma N x}{2 \left(E^b\right)^2} \left(1+2x \frac{\partial_x N}{N}\right)- \frac{\gamma N}{2 x} - \frac{N}{2 \gamma \Delta x} \partial_x \left(\frac{x^3}{\Delta} \sin\left(\frac{\sqrt{\Delta} b}{x}\right)^2\right) \;,\label{eq:dotbfull}
	\end{align}
\end{subequations}
\noindent
but still have a very similar form.

We are interested in static solutions of these equations i.e., when $\dot{E}^b = \dot{b} = 0$.
The first condition, $\dot{E}^b = 0$ is due to Eq.~\eqref{eq:dotEbfull} satisfied for either
\begin{equation}
	N^x = 0 \quad \text{or} \quad \partial_x \left(\frac{N E^b}{x}\right) = 0 \;,
\end{equation}
\noindent
which gives two classes of possible solutions.
As it turns out the class of $N^x = 0$ is contained in the second one and thus it is not considered here.
The second equation implies
\begin{equation}\label{eq:Ebfull}
	\frac{E^b}{x} = \frac{D}{N} \;,
\end{equation}
\noindent
where $D$ is an arbitrary integration constant which can be fixed later on.
Demanding that $b$ also be static leads, according to Eq.~\eqref{eq:dotbfull}, to
\begin{align}
	0 =& - \frac{x^2 \partial_x E^b}{\left(E^b\right)^3} + \frac{3 x}{2 \left(E^b\right)^2} -\frac{1}{2x} - \frac{1}{2 \gamma^2 \Delta x} \partial_x\left(x^3 \sin\left(\frac{\sqrt{\Delta} b}{x}\right)^2\right) 
	\notag
	\\
	=& \frac{1}{2x} \partial_x\left(\frac{x^3}{\left(E^b\right)^2}\right) -\frac{1}{2x} - \frac{1}{2 \gamma^2 \Delta x} \partial_x\left(x^3 \sin\left(\frac{\sqrt{\Delta} b}{x}\right)^2\right) \,.
\end{align}
\noindent
This can easily be integrated with the result
\begin{equation}\label{eq:sinbfull}
	x - C = \frac{x^3}{\left(E^b\right)^2} - \frac{x^3}{\gamma^2 \Delta} \sin\left(\frac{\sqrt{\Delta} b}{x}\right)^2 \quad \Rightarrow \quad \sin\left(\frac{\sqrt{\Delta} b}{x}\right)^2 = \frac{\gamma^2 \Delta}{x^2}\left(\frac{N^2}{D^2} - 1 + \frac{C}{x}\right) \;,
\end{equation}
\noindent
where Eq.~\eqref{eq:Ebfull} was used in the last step.
This leads immediately to the shift, given by
\begin{equation}\label{eq:Nxfullsol}
	N^x = -N \sqrt{\left(\frac{N^2}{D^2} - 1 + \frac{C}{x}\right)\left(1-\frac{\gamma^2 \Delta}{x^2}\left(\frac{N^2}{D^2} - 1 + \frac{C}{x}\right)\right)} \;.
\end{equation}
\noindent
Note that here we have a sign choice as the root of Eq.~\eqref{eq:sinbfull} has to be taken.
This sign choice was absorbed in $N$.
Exactly this sign is important to construct the continuous extension in the main text.
The Hamiltonian constraint $\H^{eff} = 0$ is trivially satisfied as it can be rewritten as a derivative of Eq.~\eqref{eq:sinbfull}.
Therefore, the spacetime is fully determined up to the free gauge choice $N$.
The general metric then reads

\begin{equation}\label{eq:metricfull}
	\dd s^2 = - N^2\left(1-\frac{\left(N^x\right)^2}{D^2}\right) \dd t^2 +2 \frac{N^x D^2}{N^2} \dd t \dd x + \frac{D^2}{N^2} \dd x^2 + x^2 \dd \Omega^2\;,
\end{equation}
\noindent
with $N^x$ given by Eq.~\eqref{eq:Nxfullsol} and only dependent on the lapse $N$ and the integration constants $D$ and $C$.

Note that it is possible to choose another time coordinate $\tau = t/D$, which makes the metric only dependent on the combination $N/D$.
Thus, also redefining the lapse $\bar{N} = N/D$ makes the line element independent of $D$.
The freedom in choosing $D$ is thus absorbed in the freedom to choose $N$.
This family of solutions, once the function $\bar{N}$ is fixed, depends on only one integration constant, which can be related to the mass.
Its precise dependence on the mass might depend on the choice of $N$.
This fits perfectly into the picture of \cite{BodendorferMassandHorizon}, where the number of Dirac observables i.e., integration constants with physical effects for several black hole models were studied.
The main difference is that the setting of \cite{BodendorferMassandHorizon} is static from the beginning.
As argued there performing a polymerisation, which is independent under fiducial cell rescalings (only present in the static case), leads to a quantum theory with only one free parameter i.e., the black hole mass.
Indeed, in the models \cite{KellyBlackholecollapse,KellyEffectiveloopquantumgravity} a polymerisation of $b$ takes place, which is in the static case independent of the fiducial cell.
This comparison is very vague as the basic underlying formulations are very different, but in this sense, the observation that Eq.~\eqref{eq:metricfull} only depends on $C$ perfectly agrees with the results in \cite{BodendorferMassandHorizon}.

It is also easily possible to rediscover the Planck-scale braking of diffeomorphism invariance noted in \cite{KellyEffectiveloopquantumgravity} from this point of view.
Computing spacetime scalars, e.g. $R_{\mu \nu \alpha \beta} R^{\mu \nu \alpha \beta}$ should not depend on the lapse $N$ as this is usually a pure gauge degree of freedom.
Nevertheless, performing this computation for Eq.~\eqref{eq:metricfull} gives terms proportional to $N$ and its first and second derivatives (for $R_{\mu \nu \alpha \beta} R^{\mu \nu \alpha \beta}$), which are all suppressed by factors of $\Delta$.
As spacetime scalars depend now on $N$ i.e., the specific time coordinate the spacetime is only diffeomorphism invariant up to corrections of order $\mathcal{O}\left(\Delta\right)$.
This certainly has to be understood better in the future, but it seems so far to be a feature (or bug) of most of the LQG-inspired black hole models \cite{BojowaldAno-goresult}.
A possible interpretation of this in the context of the quantum theory was provided in \cite{GambiniLQBHextensions}.
The authors conclude that the breaking of diffeomorphisms can be related to the quantum discreteness of spacetime and different realisable observers, whose observations coincide in the regime where the discreteness can be neglected.

It would be interesting to view $N$ not as a lapse but rather as a Dirac observable, which has to be fixed by suitable initial or boundary conditions and physical input.
To address this problem, it would be important to work out the physical role of $N$ with the most general solution \eqref{eq:metricfull}.
Reversely, it might also be that the presence of a second Dirac observable in \cite{BodendorferMassandHorizon} indicates a dependence on the lapse if embedded to a dynamical theory.
This second observable would then indicate the non-covariance of these polymer models, which would be consistent with \cite{BojowaldAno-goresult}.

Obviously, the results of \cite{KellyEffectiveloopquantumgravity} can be reproduced.
For $\bar{N}=N/D = 1$ and $C = R_s$ the vacuum metric Eq.~\eqref{eq:dsext} is recovered.
Also, the metric like that of \cite{GambiniSphericallysymmetricloop}, discussed in \cite{KellyEffectiveloopquantumgravity} can be reproduced by choosing $\bar{N} = N/D = 1/\sqrt{1+R_s/x}$ and $C  = R_s$.
These are now simply special cases of the more general metric \eqref{eq:metricfull}.
Particularly important is the case where $\bar{N}=N/D = -1$ and $C = R_s$, which is exactly the vacuum metric of the main text only time-reversed.
This metric plays a crucial role in Sec.~\ref{sec:discont} to remove the discontinuity.
The sign choice of $N_{vac}^x$ in Eq.~\eqref{eq:Nxvacsign} corresponds then simply to the choice $\text{sign}(N) = \pm1$.


\end{document}